\documentclass[epj-spec]{svjour}
\usepackage{graphics}
\usepackage{epsfig}
     
     \newcommand{\pathnow}{}
\def\lessim{\lower.5ex\hbox{$\; \buildrel < \over \sim \;$}}
\def\gtrsim{\lower.5ex\hbox{$\; \buildrel > \over \sim \;$}}
\newcommand{\nc}{\newcommand}
\nc{\ds}{\displaystyle}        \nc{\ts}{\textstyle}
\nc{\rf}[1]{Fig.\,\ref{#1}}    \nc{\rt}[1]{table\,\ref{#1}}
\nc{\req}[1]{Eq.\,(\ref{#1})}  \nc{\eps}{\varepsilon}
\nc{\beq}{\begin{equation}}     \nc{\beql}[1]{\begin{equation}\label{#1}}
\nc{\eeq}{\end{equation}}        
\nc{\beqa}{\begin{eqnarray}}   \nc{\eeqa}{\end{eqnarray}}       
\nc{\bfi}{\begin{figure}}       \nc{\efi}{\end{figure}}

\begin{document}
\title{Strangeness Enhancement}
\subtitle{Challenges and Successes}
\author{Johann Rafelski\thanks{\email{rafelski@physics.arizona.edu}\\ Dedicated to memory of  
Prof. J\'ozsef Zim\'anyi} }
\institute{Department of Physics, University of Arizona, Tucson, AZ 85721}
\abstract{Highly effective conversion of kinetic energy into abundant particle multiplicity is the remarkable 
feature discovered in high energy heavy ion collisions. This short and pedagogic review addresses 
topical issues related to the understanding of this phenomenon, originating in the creation of the 
deconfined quark--gluon plasma phase. I consider in depth  the apparently simple, yet
  sometimes misunderstood, intricate  issues: 
a) statistical hadro-chemistry, chemical parameters,  
b)  strange flavor chemical equilibration in quark--gluon plasma, and 
c) particle yields and sudden hadronization, 
in the historic perspective of work and competition with  my friend J\'ozsef Zim\'anyi.
} 
\maketitle
%

\section{Creation of Matter in Laboratory} \label{matter}
Today, we are at the verge of understanding  the origin of matter surrounding us.
 How did visible matter which surrounds us, all
the protons and neutrons, i.e., `hadrons' form, emerging 
from the quark--gluon soup which filled the early Universe? 
In  laboratory experiments involving collisions of large nuclei at relativistic energies,
several (nearly)   independent reaction steps occur, and ultimately are   leading to  hadron production:\\
\indent 1) formation of  the primary fireball; a momentum equipartitioned partonic 
phase comprising in a limited space-time domain the final state entropy;\\Z
\indent 2) the cooking of the energy content of the hot matter fireball
  towards the yield (chemical) equilibrium in a hot perturbative quark--gluon 
plasma phase --- this is the quark--gluon plasma liquid (QGP) --- a drop of the matter that filled the 
universe up to about 30$\mu$s;\\
\indent 3) emergence near to the phase boundary of  {\em transient}  massive effective quarks and 
disappearance of free gluons; this phase cannot be in chemical equilibrium if entropy, energy, 
baryon number and strangeness are to be conserved;\\
\indent 4) hadronization, that is combination of  effective and strongly interacting $u,d,s, \bar u,\bar d$ and $ \bar s$ quarks 
and   anti-quarks   into the final state  hadrons, with  the yield probability weighted by accessible phase space. 

In particular, the hadronization process can be subject to detailed experimental study, and one of the 
points of interest is how, as function of increasing energy of colliding nuclei, the hadronization of the dense 
quark--gluon matter fireball  occurs --- we argue 
that at sufficiently high energy, e.g., at RHIC,  this happens 
 in a rapid and explosive manner,  forcing the global  bulk matter into `single'
time freeze-out `hadronization'.
The methods we discuss here should also allow to settle   open questions about  
low energy reactions,  i.e., how fast 
hadronization proceeds at  SPS range of energies and if  hadrons remain
hadrons  without quark degrees of freedom appearing in the reaction at all, or perhaps there is 
an intermediate domain of constituent quark matter. This is the `low energy' heavy ion frontier. The high energy 
frontier is the arrival of LHC-CERN heavy ion research program which will allow us to explore the 
extreme conditions of perturbative quark--gluon plasma. In these notes, we will focus on RHIC and LHC 
case only. 

The abundant  production, in these `high energy'  heavy ions reactions, 
 of  strange flavored hadrons is a direct consequence of the 
process of chemical equilibration of strange quarks in QGP. 
In the Summer 1980, I proposed that the strangeness abundance, along with
strange antibaryon  yields offers an opportunity  to identify formation of the 
deconfined quark--gluon matter, and the exploration of  its properties.
The original argument as stated at the time follows in the following 
verbatim \cite{Rafelski:1980rk}, since I could
not really present it better today\footnote{Another   reason for the verbatim presentation of 
this  text is  that no regular publication occurred,
and libraries  tend to dispose of the old 1980 conference volumes --- for this reason 
a scan of this articles is  also available at http://www.physics.arizona.edu/\~\,rafelski/rare.htm:}:

{\it \ldots assuming equilibrium in the quark plasma,
we find the density of the strange quarks to be (two spins and three colors)\,\footnote{I changed here  the notation
$s\to N_s,\bar s \to N_{\bar s}$ etc, to correspond to the  notation of this presentation.}:
\begin{equation}\label{Eq1}
\frac{ N_s} V =\frac{ N_{\bar s}} V =6\,\int \frac{d^3p}{(2\pi)^3}e^{-\sqrt{p^2+m_s^2}/T}=3\frac{Tm_s^2}{\pi^2}K_2(m_s/T),
\end{equation}
(neglecting, for the time being, the perturbative corrections and, of course, ignoring weak decays). As the mass of the
strange quarks, $m_s$, in the perturbative vacuum is believed to be of the order of 280--300 MeV\footnote{This high value
of strange quark mass based on qualitative consideration of hadron spectra,  was soon recognized to be a factor two too 
large, and further 
underwent a slow but steady reduction as function of time and the development of better analysis
methods. Today, value is just about 1/3 as large, an thus assumption of equilibrium in QGP is less of a challenge than it 
seemed in the year 1980.}, the assumption of equilibrium for $m_s/T\simeq 2$ indeed correct. In Eq.\,(\ref{Eq1}), we were
able to use Boltzmann distribution, as the density of strangeness is relatively low. Similarly, there is a certain light 
anti-quark density ($\bar q$ stands for either $\bar u$ or $\bar d$):
\begin{equation}\label{Eq2}
\frac{N_{\bar q}}{V}\simeq 6\int  \frac{d^3p}{(2\pi)^3}e^{-|p|/T-\mu_q/T}=e^{-\mu_q/T} T^3 \frac{6}{\pi^2},
\end{equation}
where the quark chemical potential is $\mu_q=\mu_B/3$, $\mu_B$ is baryochemical potential. This exponent 
suppresses the $q\bar q$ pair production as only for energies higher than $\mu_q$ is there a large number of empty 
states available for the $q$. 

What we intend to show is that there are many more $\bar s$ quarks than anti-quarks of each light flavor.  Indeed:
\begin{equation}\label{Eq3}
\frac{N_{\bar s}}{N_{\bar q}}=\frac 1 2 \left( \frac{ m_s}{T}\right)^2K_2(m_s/T)e^{\mu_B/(3T)}.
\end{equation}
The function $x^2K_2(x)$ is, for example, tabulated in Abramowitz-Stegun. For $x=m_s/T$ between 1.5 and 2, it
varies between 1.3 and 1. Thus, we almost always have more $\bar s$  than $\bar q$ quarks and, in many cases
of interest $N_{\bar s}/N_{\bar q}\simeq 5$. As $\mu_B\to 0$ there are about as many $\bar u$ and $\bar d$ quarks as there
are $\bar s$ quarks. 

When the quark matter dissociates into hadrons, some of numerous $\bar s$ may, instead of being bound in a 
$q\bar s$ Kaon, enter into a ($\bar q \bar q \bar s$) anti-baryon \ldots}

There are three important issues raised above, which since have seen a long and tedious further development:
\begin{enumerate}
\item  the 
chemical equilibration of strange quarks;
\item the combinant quark
hadronization;
\item the importance of strange  anti-baryons as signature of QGP. 
\end{enumerate}
  Very few people at first noted
our work, let alone evaluate its relevance and validity. 
However, following several lectures, in particular also at the October 1980  GSI
workshop\footnote{Workshop organized by R. Bock and R. Stock
on {\it Future Relativistic Heavy Ion Experiments} held at GSI-Darmstadt, October 7--10, 1980, in GSI 81-6 report.}
which was attended by  J\'ozsef Zim\'anyi, already 
by the Summer 1981, Tam\'as   B\'{\i}r\'o working with J\'ozsef Zim\'anyi obtained
strangeness production rates in perturbative QCD when considering the specific
 processes $q\bar q\to s\bar s$. The result of this study was that it would take much
too long, about 8 times  the natural lifespan of a QGP fireball, to equilibrate strangeness chemically.
Unfortunately, when  J\'ozsef Zim\'anyi came to present these yet preliminary results 
in Frankfurt in late Summer 1981, I was not aware of this visit  set-up on a   short notice  by my Institute 
director, Walter Greiner. Upon my return from Seattle,  Walter greeted me with 
the announcement, ``Johann, J\'ozsef Zim\'anyi has shown your strange  theory could be wrong". For a
young man at the beginning of his carrier this was quite an event, hard to forget. 

Walter   has been  convinced by the presentation, moreover, he needed my talents
for the study of the vacuum in  strong fields. As weeks passed, the pressure in Frankfurt increased: 
Walter argued that I  should  cease all `strange' quark--gluon 
activity which,   as   J\'ozsef Zim\'anyi has shown, could not be right. Finally, 
 I believe   in late October 1981, I (and Walter\footnote{Walter's copy was in
 my mailbox marked: $\Rightarrow$ Rafelski: {\it Johann, bitte R\"ucksprache dar\"uber! Walter}}) received  a copy of
 the  Bir\'o--Zim\'anyi preprint \cite{Biro:1981zi}. For the first time, I saw in any 
detail the contents of J\'ozsef's  lecture, and I saw an 
important omission. 

Among my CERN-earned credentials 1977--80 was some  knowledge about charm production
in $pp$ reactions: I shared, for about a year, an office with Brian Combridge, of 
perturbative QCD charm production fame \cite{Combridge:1978kx}.
I was induced to Hagedorns
relativistic thermodynamics and statistical bootstrap, and in parallel  also, 
to perturbative QCD charm production. In the fall 1981, this  turned out to be 
a valuable asset. I had learned from Brian that even if the cross sections were similar  for both 
quark $q\bar q\to c\bar c$ and gluon  $GG\to c\bar c$ fusion processes into charm, it was
the gluon which dominated the production  rate. Once I saw that   $GG\to s\bar s$ 
was not considered in the  B\'{\i}r\'o--Zim\'anyi  preprint, I knew that the essence of the calculation
of the strangeness production relaxation time remained open. I described  my insight to Walter, who 
did not like the idea that  gluons, at the time very hypothetical objects, would be in essence 
the cause of strangeness production (a speculative signature)
in the quark--gluon plasma, another highly hypothetical 
object. On the other hand, my   colleague Berndt M\"uller was enthusiastic at the prospect to
use real gluons in a physical process. We had just  completed a study based on virtual 
gluon fluctuations of the  temperature dependence
of the latent heat of the QCD Vacuum \cite{Muller:1980kf}, and thus the glue based 
flavor producing reactions did not seem exotic at all given this preparation.

 Within  a few days of work,  we found  that indeed  the thermal
 strangeness chemical equilibrium in quark--gluon 
plasma is due to gluon fusion process \cite{Rafelski:1982pu}.  
The issue of chemical equilibration in QGP  became an asset, for QGP chemical equilibrium 
yield of strangeness evolved into an indicator of the presence of mobile, free gluons,
and thus of deconfinement: The work of  B\'{\i}r\'o--Zim\'anyi showed that as long
as there was no free glue, just light $u,\bar u, d, \bar d$ 
quarks, chemical equilibration was not attainable.   This was 
elaborated for the physical case of a hadron gas  in a kinetic approach by  another 
talented student, Peter Koch. He   computed the strangeness  yields and relaxation times 
expected in  the hadron phase \cite{Koch:1984tz,Koch:1986ud}.
By 1986, Strangeness yield enhancement has been well established theoretically 
as signature  for deconfinement,
and the strength of this enhancement was recognized to depend on how long 
the hot QGP phase would last.

An important aspect in the evaluation of the rate of strangeness production, which 
Berndt and I undertook, was the choice 
of the value of the running   strong coupling constant $\alpha_s(\Lambda)$.
We knew very well  that if one uses a 1st-order 
perturbative expression for a QCD process, it can only produce reasonable results 
if the coupling strength is  chosen at the right strength for the energy scale.  
We studied the  results available and decided that the right  $\alpha_s$
value to produce strangeness at the typical thermal collision energy, $\Lambda\simeq 3$--$6T$,
for $T=200$--$300$ MeV should be $\langle\alpha_s\rangle=0.6$. This turned out to be  just the right choice,
as later precision measurements of  $\alpha_s$ have shown\footnote{The particle data group 
 links to a web page where one can 
enter the reaction energy scale and gets back the strength of the QCD coupling,
see http://www-theory.lbl.gov/\~\ ianh/alpha/alpha.html, the value  $\alpha_s(0.86\rm{GeV})=0.60\pm0.10\pm0.07$ 
is obtained.}.   Yet, for the following 15 
years, a  value  $\langle\alpha_s\rangle=0.2$  was often used  in literature,
 and our results were not, in general, trusted --- however, this 
smaller value is  
appropriate for the energy scale $\Lambda\simeq  6$ GeV,   beyond the energy scales  
 where thermal strangeness production is occurring.

It is important to recognize  that reaction rates grow 
with  $\alpha_s^2$, hence  using $\langle\alpha_s\rangle=0.2$ instead 
of $\langle\alpha_s\rangle=0.6$ one finds a relaxation time which  is longer  by an order of magnitude.
While we predicted $\tau\simeq 2$fm/c,  I saw also results as large as  15 fm/c. An  uninvolved observer 
could not readily recognize  that this was simply a 
consequence of a different and seen from historic perspective
wrong choice of the QCD interaction strength. Indeed,  I 
even had the impression that the authors of these papers did not see that their results
differed mostly and nearly only due to different choice of the coupling strength.
About the same time ,  $\alpha_s^2$ was measured precisely, and first 
systematic study of strangeness enhancement  appeared;
a rapid shift of opinion ensued following the Warsaw ICHEP\footnote{See proceedings of the 
28th International  Conference on High Energy Physics, July 25 -- 31, 1996, 
Z. Ajduk and A.K. Wr\'oblewski, Editors,  World Scientific, Singapore.} conference  in 1996. 
Despite this, still  today, more than 10 years later  some confusion about strangeness 
chemical equilibration  lingers on. To set the record straight, let us clearly state that: \\
\indent a) at RHIC-200 it is the QGP phase which equilibrates chemically, \\
\indent  b) produced hadron abundances are determined by the QGP hadronization process.\\
It is important to note that these results are independently a consequence of both, theoretical analysis based on
known fundamental properties of strong interactions, and an analysis of experimental data.  Considering the totality 
of RHIC results pointing to rapid, nearly instantaneous hadronization, and early parton thermalization, I see no 
path leading to the  chemical equilibrium among hadrons produced. The remainder
of this paper will lay foundation for understanding of this situation.

\section{Statistical Hadronization} \label{stathad}
Since the beginning of the friendly science competition  I have had with 
J\'ozsef Zim\'anyi and his group, much of our effort was devoted to the
understanding of how exactly production of hadrons occurs, and 
the methods we developed are similar, with the overriding element 
being the conservation of energy, flavor and consideration of entropy
so that it cannot decrease in the recombinant mechanisms. 

The Budapest group obtained the hadron multiplicities within the 
framework of the algebraic coalescence re-hadronization model 
(ALCOR) \cite{Biro:1994mp}, not much different of my efforts. However,
my focus at the time was to develop methods of diagnosis of the QGP 
using specifically strange particle production \cite{Koch:1986ud,Rafelski:1991rh}, 
from which emerged the 
Fermi-2000 hadron production model (for a review, see \cite{Letessier:2000ay}).
An elaboration of this effort is the SHARE (Statistical HAdronization with REsonances)
suite of programs \cite{Torrieri:2004zz,Torrieri:2006xi}.
We find that the  overall hadronic particle yields  are 
well described by the   statistical hadronization model  (SHM),
a point made by several other groups as well \cite{Becattini:2001fg},
including features of the hadron spectra. 

The  SHM is an elaboration of the Fermi model of hadron formation,
based on the hypothesis
 that the strong interactions  saturate the   quantum particle production
matrix elements. Therefore,  the  yield  of   particles  is controlled primarily by the 
accessible phase space, and not by reaction strength. 
In the original Fermi model, the accessible 
phase space is considered in terms of the available energy. We refer to this
approach  as `micro-canonical'. Given the large energy contents of the fireball
 we use the (grand) canonical approach, with
a  temperature-like parameter $T$. The SHM contains little if any 
information about the nature of interactions, and thus it embodies the 
principle of reaching simplicity in many body dynamics, allowing to 
identify the properties of the dense and hot primary matter formed in 
heavy ion collisions. 

\subsection{Chemistry parameters explained}
We now consider either the phase space into which particles are emitted, or, possibly 
the actual properties of the hot matter, such as QGP, or even hadron gas, that is a phase of 
interacting confined hadrons.  The objective is to understand how to describe the yields
of the multitude of different hadrons emerging.

Each of the hadrons produced will have some quantum numbers
such as baryon number, strangeness, etc. which need to be followed
and conserved in the reactions.
  An  direct way to accomplish
this   objective consists in  characterizing 
each particle by the  valance quark content and the related
 quantum numbers  \cite{Koch:1982ij},
we shall establish the relation between valance
quark count and the hadron properties below. 

In principle,  the yield of each particle is  governed,  aside of
other statistical parameters such as the size of the system (volume $V$)  and 
the (chemical) freeze-out temperature $T$, by the particle fugacity obtained,
as we shall argue, 
forming a product of two types of chemical factors,
\begin{equation}\label{ups}
\Upsilon_i\equiv \lambda_i^{\pm 1}\gamma_i= e^{ \sigma_i^\pm  /T},
\end{equation}
where $ \sigma_i^\pm$
 is  `$i$'-particle (antiparticle for lower sign) chemical potential $\sigma_i^{\pm}$. We see that for each related  
particle and antiparticle,  we   have a different   value of $\Upsilon_i$ and thus, if we wish so,
of particle  chemical potential ($=T\ln \Upsilon_i$). Our approach, i.e., use
of $\gamma_i$ and $\lambda_i$ is   convenient for  the control of
the difference, and the sum of particles and antiparticles separately 
as we shall explain just below. Our way of handling chemistry  is in principle  not
different from the ALCOR approach  \cite{Biro:1994mp},  
in which  two independent chemical
 potentials for  quarks and antiquarks are used. For two reasons, we believe 
that this way is better; the book-keeping of multitude of particles is easier, 
and the interpretation of the factors in a microscopic model has the
advantage of relating them directly to different type of microscopic reactions. 

Thus, in our approach, there   are  two types
 of chemical factors $\gamma_i$ and $\lambda_i$. These factors are related to  
two types of chemical equilibrium, which can be better understood
 when one introduces $\gamma_i,\lambda_i$ as
defined above. These are described  in table~\ref{parameters}. 
Moreover, there is considerable difference in the dynamics of the (chemical) 
reaction associated
with these parameters, even though, in principle, there is 
no difference in their fundamental origin and function.  To understand
better this somewhat special situation, let us focus on  
 strangeness in the hadronic gas phase.  
The two principal chemical processes are seen in \rf{exchange}.
The  redistribution of strangeness among 
(in this example) $\Lambda,\,\pi$ and $N,$ K  seen on left in \rf{exchange},
constitutes approach to the `relative'  chemical equilibrium 
of these species. In this equilibration process, the available quarks 
are distributed among hadrons. The picture refers to a typical reaction
in which, e.g., $\Lambda +\pi\leftrightarrow N+K$. The $s,\bar s$ pair
 production process, on right in \rf{exchange}, 
is responsible for absolute   
chemical equilibrium of strangeness. Achievement of the 
absolute equilibrium, $\gamma\to 1$, require  more rarely 
occurring  collisions with annihilation and creation of new (strange) particle
pairs, e.g., $N+\pi \leftrightarrow \Lambda+K$. 
These reactions are `OZI' forbidden, which means that the cross sections are 
about a factor 10--20 weaker compared to the exchange processes, 
and thus they are  slower in driving  the `absolute' chemical 
equilibrium.

\begin{table}[hbt]
\caption{\label{parameters}Four quarks $s,\ \overline{s},\ q,\ \overline{q} $
 require four chemical parameters; right: name of the associated chemical equilibrium.} 
\vskip 0.3cm
\begin{center}
\begin{tabular}{ll|l}
\hline
\hline 
  $\lambda_{i}$&   controls  `nett'   i.e. difference yield & Relative   chemical\\
&quarks$_i$ $-$ anti-quarks$_i$  ($i=q,s$) & equilibrium\\
\hline
$\gamma_{i}$&   controls overall abundance & Absolute   chemical\\
&of quark$_i$  ($i=q,s$)  pairs & equilibrium 
\end{tabular}
\end{center}
\end{table}

\begin{figure}[tb] 
\begin{center}
\psfig{width=5cm,figure=\pathnow 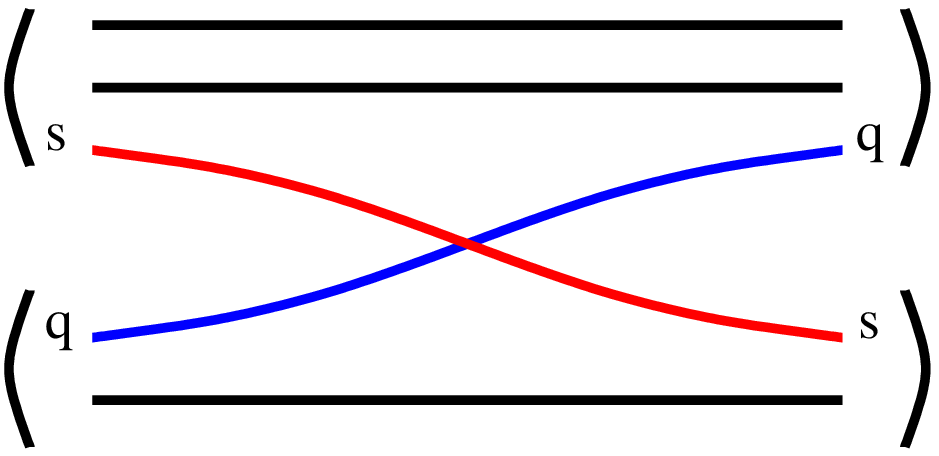}
\hspace*{1cm}\psfig{width=5cm,figure=\pathnow 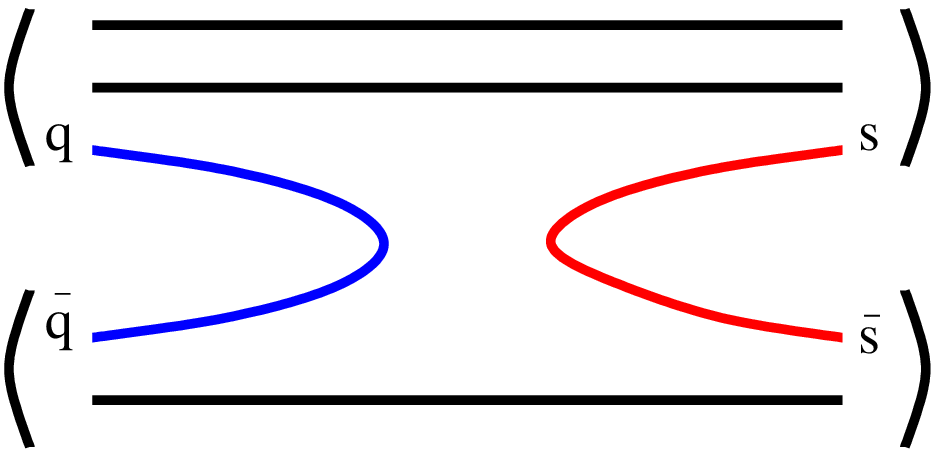}
\end{center}
\caption{\label{exchange}
Typical strangeness exchange (left) and production (right)
reactions in the hadronic gas phase.}
\end{figure}

As a first example, let us illustrate why $\lambda$ controls the difference between
 particle and antiparticle number, and  $\gamma$ counts in effect the yield of  particle 
and antiparticle pairs. To see this, let us look  at  the simple but artificial case of a 
gas of  nucleons and anti-nucleons $N,\overline{N}$. The
two chemical fugacities are:
\begin{equation}
\Upsilon_N=\gamma_N e^{ \mu_N /T}, \qquad\qquad
\Upsilon_{\overline{N}}=\gamma_N  e^{- \mu_N /T}.
\end{equation}
Thus, we find for the potentials:
\begin{equation}
\sigma_{N}\equiv \mu_N+T\ln\gamma_N ,\qquad
\sigma_{\overline{N}}\equiv -\mu_N +T\ln\gamma_N.
\end{equation}
Considering the first  law of thermodynamics: 
\begin{eqnarray}
dE+P\,dV-T\,dS&=&\sigma_N\,dN+
      \sigma_{\overline{N}}\,d\overline{N},\\ \nonumber
&=&
 \mu_N (dN-d\overline{N})+ T\ln \gamma_N (dN+d\overline{N}),
\end{eqnarray}
we recognize that the nucleon chemical potential  
 $\mu_N$ controls the net nucleon  number  arising from the particle difference, while
$\gamma_N$, which we call the phase space occupancy (of nucleons),  
 regulates the number of nucleon--antinucleon pairs present. 

In the next example, we consider how we use the quark parameters when dealing with hadrons. 
Noting  the valance quark content, of   $p(uud),\bar p(\bar u\bar u\bar d)$, we have 
\begin{equation}
\Upsilon_{p}=(\gamma_u^2\gamma_d)\ (\lambda_{u}^2\lambda_{d})
 \qquad\qquad
\Upsilon_{\bar p}=(\gamma_u^2\gamma_d)\ (\lambda_{u}^{-2}\lambda_{d}^{-1}),
\end{equation} 
and thus we can write:
\begin{equation}
\Upsilon_{p} 
={\gamma_u^2\gamma_d}\,e^{2\mu_u+\mu_d\over T} 
={\gamma_B}\,e^{\mu_B\over T},
\qquad
\Upsilon_{\bar p} 
={\gamma_u^2\gamma_d}\,e^{-2\mu_u-\mu_d\over T} 
  ={\gamma_B}\,e^{-\mu_B\over T}.
\end{equation} 
In case of   $\Lambda(uds),\overline\Lambda(\bar u \bar d\bar s)$, we need to respect
an  important  historical   anomaly, i.e.,  negative 
S-strangeness assignment to  $s$-hadrons, {e.g.}:
\begin{equation}
\Upsilon_\Lambda=\gamma_u\gamma_d\gamma_s\,e^{\mu_u+\mu_d+\mu_s\over T}
           ={\gamma_B\over \gamma_{\mathrm{S}}}\,e^{\mu_B-\mu_{\rm S}\over T},
\quad
\Upsilon_{\overline\Lambda}=\gamma_u\gamma_d\gamma_s\,e^{-\mu_u-\mu_d-\mu_s\over T}
   ={\gamma_B\over \gamma_{\mathrm{S}}}\,e^{-\mu_B+\mu_{\rm S}\over T}.
\end{equation}
This anomaly was the historical reason why we (I and all my collaborators) 
in general follow quark flavor and use quark chemical factors
to minimize the confusion arising. There is a second reason, as we shall see just below.  First,  
in order to focus on the important parameters,
and considering  the good symmetry between $u$ and $d$ quarks, we simplify notation,
allowing us to refer generically to the light quarks $q$:
\begin{equation}
 \lambda_q\equiv \sqrt{\lambda_u\lambda_d},\quad \mu_q={\mu_u+\mu_d\over 2};\qquad 
 \lambda_{I3}\equiv \sqrt{\lambda_u\over \lambda_d},\quad \mu_{I3}={\mu_u-\mu_d\over 2}.
\end{equation}
 and similarly for $\gamma_q$ and $ \gamma_{I3}$.

We thus now can write for the above examples of $p$ which we call generically $N(qqq)$ for nucleon,
$\Lambda$ which we call generically $Y(qqs)$, adding further the case of double-strange
$\Xi(qss)$:
\begin{eqnarray}
&\Upsilon_{N} 
={\gamma_q^3}\,e^{3\mu_q\over T} 
={\gamma_B}\,e^{\mu_B\over T},
&\quad
\Upsilon_{\overline N} 
={\gamma_q^3}\,e^{-3\mu_q\over T} 
  ={\gamma_B}\,e^{-\mu_B\over T},
\nonumber\\[0.2cm]
&\Upsilon_Y=\gamma_q^2\gamma_s\,e^{2\mu_q+\mu_s\over T}
           ={\gamma_B\over \gamma_{\mathrm{S}}}\,e^{\mu_B-\mu_{\rm S}\over T},
&\quad
\Upsilon_{\overline Y}=\gamma_q^2\gamma_s\,e^{-2\mu_q-\mu_s\over T}
   ={\gamma_B\over \gamma_{\mathrm{S}}}\,e^{-\mu_B+\mu_{\rm S}\over T},
\nonumber\\[0.2cm]
&\Upsilon_\Xi=\gamma_q\gamma_s^2\,e^{\mu_q+2\mu_s\over T}
           ={\gamma_B\over \gamma_{\mathrm{S}}^2}\,e^{\mu_B-2\mu_{\rm S}\over T},
&\quad
\Upsilon_{\overline\Xi}=\gamma_q\gamma_s^2\,e^{-\mu_q-2\mu_s\over T}
   ={\gamma_B\over \gamma_{\mathrm{S}}^2}\,e^{-\mu_B+2\mu_{\rm S}\over T}.
\end{eqnarray} 
 Naturally, I can write similar equations  
  for the $\Omega$ and $\overline \Omega$, kaons, and all strange particles in general.

 The second historical anomaly is that we recognize in above relations that,  in order to   
relate the quark and hadron quantum numbers, there is only one choice:
\begin{equation}
\mu_B=3\mu_q,\  \mu_{\mathrm{S}}=\mu_q-\mu_s; \qquad
 \mu_q=\frac 1 3 \mu_B, \ \mu_s= \frac 1 3 \mu_B-\mu_{\mathrm{S}} .
\end{equation}
To make the 2nd anomaly explicit: these relations seem to imply that
 there is no baryon number in the strange quarks! $\mu_s$ does  not enter into the 
magnitude of $\mu_B$. One can trace this anomaly to the assignment of the full baryon number to $Y$, $\Xi$ 
along with `strangeness' $S=-1$ and $-2$ quantum numbers respectively. 
Thus, even though strange quarks contain baryon number
they are counted as being `baryon-free'  when it comes to hadronic chemistry based on historic
definitions. 

These two anomalies have lead, as I have seen it over 25 years,  to a few   cases of confusion. Thus let me offer
instruction on  how you
can check published results without doing a complete re-computation --- this task  is easy  for baryons. One notices that:
\begin{equation}
 \left({\overline \Lambda \over \Lambda}\right)\left /  \left({\bar p\over p}\right)\right . =
 \left({\overline\Xi\over \Xi}\right)\left /  \left({\overline \Lambda \over \Lambda}\right)\right . =
 \left({\overline\Omega\over \Omega}\right)\left /  \left({\overline\Xi\over \Xi} \right)\right .=e^{+2\mu_{\mathrm{S}}/T}. 
\end{equation}
When I read a paper, I first check these relations! In fact they  must be true also for microscopic 
codes which do quark coalescence, since the particle 
specific factors cancel, and the equality must be correct to all significant figures, since a potential error can 
be confined   excited baryon states used in the evaluation of the final yields. I note that sometimes there 
is a correction,  when $\lambda_{I3}$  are allowed to 
vary from unity, in order to account for the slight up/down quark number asymmetry. In that 
case. recall to replace above $ {\bar p/p}\to \lambda_{I3}^{2}  {\bar p/p}$ and
$ {\overline\Xi/\Xi}\to \lambda_{I3}^{-2} {\overline\Xi^+/\Xi^-}$.

In the third example, consider the question, what does it mean 
to say that strangeness in QGP is in chemical equilibrium? 
In the QGP, at sufficiently high temperature, 
 the  density of strangeness flavor is described by the Fermi distributions: 
\begin{equation}\label{sdens}
\langle \frac{N_s}{V} \rangle=\langle \rho_s \rangle=\int \frac{d^3p}{(2\pi)^3} \frac 1 
   {\lambda_s^{-1}\gamma_s^{-1}e^{E(p)/T}+1}, \quad 
\langle  \frac{N_ {\bar s}}{V} \rangle=\langle \rho_{\bar s}\rangle=\int \frac{d^3p}{(2\pi)^3} \frac 1 
   {\lambda_s\gamma_s^{-1}e^{E(p)/T}+1}. 
\end{equation}
Since strangeness is produced in pairs, the number of 
strange quarks $\langle s\rangle$ is equal to that of  strange anti-quarks $\langle \bar s\rangle$, 
 and thus the QGP strangeness fugacity $\lambda_s=1$. Said differently, when local strangeness 
density is balancing, the strange quark 
chemical potential $\mu_s=T\ln \lambda=0$.  
On the other hand, the phase space occupancy parameter
$\gamma_s$ is  at the QGP formation   small, though non-zero since there
is direct production of strangeness in pre-thermal parton collisions.  Thereafter, strangeness
yield grows and at high temperature where Boltzmann approximation is very precise,
\begin{equation}
\gamma_s\simeq \frac{\rho_s+\rho_{\bar s}}{\rho_s^{\rm eq}+\rho_{\bar s}^{\rm eq}}.
\end{equation}

To recapitulate, when pair production processes are possible, the hadro-chemistry 
is greatly simplified by the introduction of the phase space occupancy 
 parameter  $\gamma_i$ which controls the abundance of pairs of particles of type $i$. 
In many reaction environments,  the absolute chemical equilibrium distribution 
$\gamma _i=1$ is reached, at which entropy is maximized. In general, the value of $\gamma_i$ 
results from dynamical development of the reaction. It is not customary   to introduce
a chemical potential associated with the occupancy fugacity $\gamma_i$ since  
the change in $\gamma_I$ is not due to chemical processes, but is solely due to particle pair 
production. On the other hand, the conventional chemical potentials describe relative 
yields of particles and reflect the properties of the medium, for example in baryon rich matter there
is an asymmetry in the yield of charged kaons, with the yield of K$^+(u\bar s)>$K$^-(\bar u d)$. 
In the deconfined phase, in order to conserve entropy
the dimensionless quantity $\mu_B/T=3\mu_q/T$ is nearly conserved in hydrodynamic
expansion of QGP. This means that $\mu_q$, $\mu_B$  evolves  in time  along with $T$.

\subsection{Entropy must increase, strangeness is conserved in hadronization}\label{entro}
The microscopic process leading to the hadronization of  QGP phase
 is not understood. However, several observables such as pion
correlations, suggest
  that hadrons are emerging within a very short time from a relatively 
small space--time volume domain \cite{Csorgo:2005gd,Florkowski:2006mb}. 
In this process, hadron formation has 
to absorb the high
entropy  content of QGP which originates in broken color bonds.
The lightest hadron is pion and most entropy per energy 
is consumed in hadronization  by producing these
particles abundantly. The particle
number, and  the entropy content follows from \req{entrof}:
\begin{equation}
{S}_\pi=
 \int\!\frac{d^3\!p\, d^3\!x}{(2\pi\hbar)^3}\,
   \left[(1+f_\pi)\ln(1+f_\pi)-   f_\pi\ln f_\pi\right]\,,
\quad 
f_\pi=\frac{1}{\gamma_{q}^{-2}e^{\sqrt{m_\pi^2+p^2}/T}-1}.
\label{newdist}
\end{equation}

As is seen in\footnote{I use this opportunity to point out that in earlier versions of this figure,
I have used a normalization appropriate for gluons, rather than pions, and  hence  
the different densities   displayed on left have considerably smaller normalization, 
while the ratios on the right are in essence unchanged.} \rf{JRABSSNE}, 
the maximum  entropy  density $S/V$ occurs for  an over-saturated  pion gas, 
$\gamma_q\simeq e^{m_\pi/2T}\simeq 1.6$. Here, 
the entropy density of such a saturated Bose  gas
is twice as large  as that of
chemically equilibrated Bose gas. Since aside of pions also
many other hadrons are produced, the large value of $\gamma_q$ 
is necessary and sufficient to allow for the smooth in $\mu_B, T$ and $ V$ 
transformation  of a QGP into hadrons. The
number of active degrees of freedom in the over-saturated hadron gas
with $\gamma_q\to \gamma_q^{\rm max}$ and in `freezing' QGP phase is
very similar.

\begin{figure}[tb]
\vspace*{1.2cm}
\centerline{
\epsfig{width=7.5cm,height=6 cm,figure=\pathnow 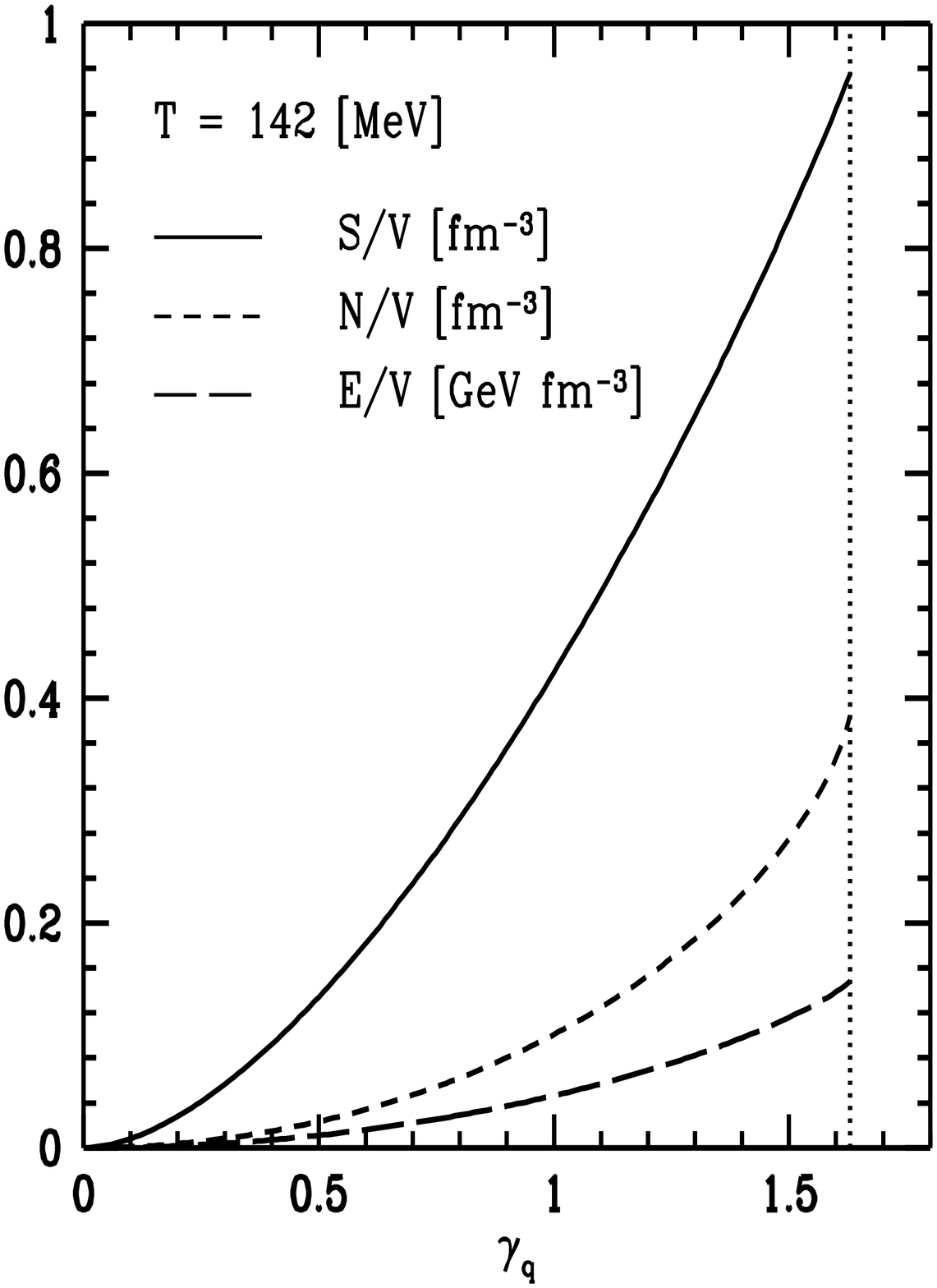}
 \hspace*{-1.cm}
 \epsfig{width=7.5cm,height=6 cm,figure=\pathnow 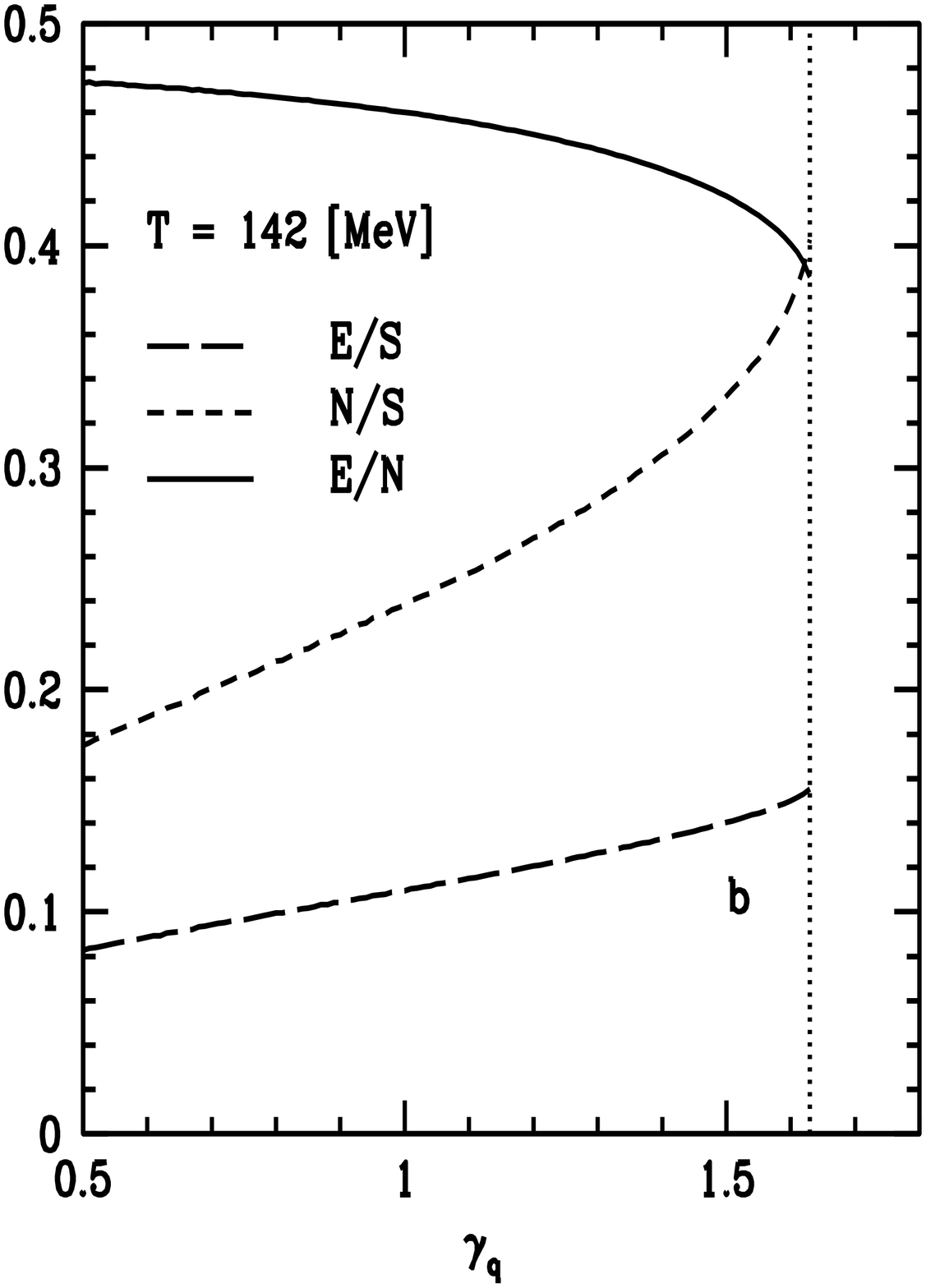}
}
\vskip -.2cm
\caption{\label{JRABSSNE}
Entropy density $S/V$ along with particle density $N/V$ and energy density 
$E/V$ as function of $\gamma_{q}$ at $T=142$\,MeV. Ratios of these
quantities are shown on right.
}
\end{figure}

We used, in \req{newdist}, a form which contains 
in an explicit way the non-equilibrium occupancy parameter. 
This can be justified noticing that 
the maximization of micro-canonical  entropy, 
\beql{entrof}
{S}_{F,B}=
 \int\!\frac{d^3\!p\, d^3\!x}{(2\pi\hbar)^3}\,
   \mp\left[(1\mp f_{F,B})\ln(1\mp f_{F,B})-   f_{F,B}\ln f_{F,B}\right]\,,
\eeq
(where indeed minus sign is for fermions and plus sign for bosons) 
subject to energy and particle (e.g., baryon, flavor, etc.)  number 
conservation  implies the quantum distributions \cite{Letessier:1993qa}:
\beql{BosFer}
 {{d^6N_i}\over{d^3pd^3x}}= {g_i\over (2\pi)^3}
{1\over  \Upsilon_i^{-1} e^{E_i/T}\pm 1},\quad
 \Upsilon_i^{\rm bosons} \le e^{m_i/T}.
\eeq
When the phase space is not densely occupied, the term $\pm 1$ in the denominator 
can be neglected and we have the commonly used  Boltzmann approximation:
\beql{Bol}
 {{d^6N_i}\over{d^3pd^3x}}=g_i{ \Upsilon_i  
\over (2\pi)^3}e^{-E_i/T}.
\eeq

In \req{BosFer}, for the Boson distribution, the value of $\Upsilon_i $ is
limited in magnitude by the Bose condensation singularity. 
This  singularity presents 
a limit on the maximum value of the fugacity, for example for pions
we have:
\beql{pimax}
\Upsilon_\pi\le e^{m_\pi/T}\equiv  (\gamma_q^{\rm max})^2\,.
\eeq
This plays a very pivotal role
considering that the mass of the pion and the hadronization temperature 
are similar. 
Large value of $\gamma_q\to e^{m_\pi/2T}$ can be directly 
noticed in pion spectra in an up-tilt in the soft portion
of the $m_\bot$ distribution. 
A similar constraint is also  arising    for $\gamma_s$
but it is much  less restrictive given the higher mass of the strangeness. Only 
in exceptional circumstance this could be of physical interest.  A more
detailed study shows that most significant constraint arises from consideration
of the $\eta$ which condenses for $\gamma_s\to 10.4$ \cite{Rafelski:2005jc}.
 
We can now compare to the entropy content of the deconfined QGP state. 
At sufficiently high temperature the entropy density $S/V$ is that of 
ideal quark-gluon gas:
\begin{equation}\label{S1}
{S\over V}={4\pi^2\over 90} g(T)T^3={\rm Const.},
\end{equation}
where we consider the quark and gluon degrees of freedom along with their
QCD corrections, extrapolating the entropy in QGP fitted to the lattice equations of state :
\begin{eqnarray}\label{ggq}
g&=&2_s8_c\left(1-\frac{ 15\alpha_s(T)}{4\pi}+\ldots\right)
+\frac74 2_s3_cn_{\rm f}  \left(1-\frac{ 50\alpha_s(T)}{21\pi}+\ldots\right).
\end{eqnarray}
The number of quark flavors is 
$n_{\rm f}\simeq 2+ \gamma_s0.5 z^2\,K_2(z)$, where $z=m_s/T$.
The   terms  proportional to chemical potentials
are not shown in the expression for entropy, since $\mu/\pi T\ll 1$ at RHIC and LHC. 

Remarkably, the different temperature dependent corrections 
cancel, and one finds that it is possible 
to use a nearly  $T$ independent value for the effective degeneracy, $g\simeq 30$ 
near to QGP breakup condition, which value is decreasing  to  $g\simeq 28$ for equilibrated 
QGP near $T=260$ MeV \cite{Kuznetsova:2006bh}. This is a relatively  large 
number comparing to  a pion gas.  However, aside of 
pions the hadron phase has many other particles and resonances and 
one finds,  in a quantitative evaluation, the value of $\gamma_q$ for 
which at a given volume (fixed in sudden hadronization) the entropy is equal
for both states. The (critical) value   $\gamma_q=1.4$  is found for $T\simeq 140$ MeV, 
decreasing with increasing temperature
and crossing  $\gamma_q=1$  at  $T\simeq 180$ MeV.

Thus, in fast hadronization of the QGP phase without an increase in volume, 
we expect that when entropy is conserved, the value of $\gamma_q$ would 
be greater than unity for every value of $T$ considered in previous studies of 
the hadronization process, with the relation being approximately:
\begin{equation}
\gamma_q\simeq1.6-0.015 (T-140)\,[\mathrm{MeV}]
\end{equation}
We are further able to evaluate the magnitude of 
$\gamma_s/\gamma_q$ after hadronization, assuming that
at RHIC and LHC the QGP phase is nearly chemically equilibrated at
point of hadronization. 
The strangeness and entropy content when carried across
the phase boundary is seen to in essence create a one-to-one 
functional relationship. For more detail, we refer to recent work of
Ms. Inga Kuznetsova \cite{Kuznetsova:2006bh}, seen in \rf{soverS}.
We recognize that at LHC where strangeness can be even over saturated, 
with the ratio of strangeness pair number to entropy reaching 0.04 (see section \ref{equQGP}),
we expect $\gamma_s/\gamma_q > 2$.
This will create an interesting challenge, which I wish  J\'ozsef Zim\'anyi 
could be part of.

\begin{figure}
\centering
\includegraphics[width=7cm,height=7cm]{\pathnow 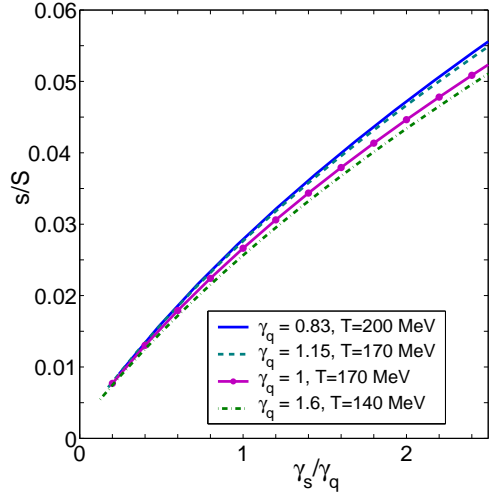}
\caption{(color on line) Strangeness to entropy ratio, $s/S$,
as a function of $\gamma_s/\gamma_q$.
 (solid line, blue) for $T=200$ MeV, $S^H=S^Q\to \gamma_q=0.083$;
(dashed line, blue) for $T=170$ MeV, $S^H=S^Q\to \gamma_q=1.15$;
 (dash-dotted line, green) for  $T=140$ MeV, $S^H=S^Q \to \gamma_q=1.6$;
(dot marked solid, violet) for $\gamma_q = 1, T=170$.}\label{soverS}
\end{figure}

\section{Strangeness production 25 years after} \label{strange}
\subsection{Strangeness per entropy}
The total final state hadron multiplicity is a  measure of the entropy $S$ produced. 
In the  QGP, the entropy  production occurs predominantly 
early on in the collision during  the parton thermalization phase.  
Once a quasi-thermal exponential energy  distribution of partons 
has been formed, the entropy production has been completed.
Since the kinetic processes leading to strangeness production are 
slower than the parton equilibration process, we are
rather certain that the production of entropy occurs mainly prior to strangeness 
production~\cite{Alam:1994sc}. However,
strangeness production by gluon fusion 
is  most effective in the early, high temperature environment,
and it continues to act during the evolution of
the   hot deconfined phase until hadronization.

Our study of  the thermal strange particle production processes 
is based on kinetic theory of particle collisions.  While the degree of chemical 
equilibration   of
gluons in early stages  (which dominate strangeness production) are uncertain, we find that, 
in particular, the observable `strangeness per entropy' $N_s/S$ (also colloquially referred to as $s/S$)
 is insensitive to this uncertainty. This insensitivity expresses the fact that at a given entropy content
the temperature can be high at low particle yield, or vice-versa, we can have large number 
of particles at low temperature.  In the cumulative strangeness  
production process, these effects compensate, as we shall here illustrate, and
the final   strangeness yield is not initial state dependent (i.e., not dependent on
full gluon equilibration). 

Both strangeness and entropy are nearly conserved near, and at
hadronization,
and thus the final state hadronic yield analysis which measures, using
the produced particle multiplicities, the value of the final state $s/S$, is closely
related to the thermal processes  in the fireball at $\tau\simeq 1$--4 fm/c. In
fact, I can  estimate the magnitude of $s/S$ in the QGP phase, considering  
the hot  early  stage of the reaction. For  an   equilibrated 
QGP phase with perturbative properties, we have discussed above:
\beql{sdivS}
{s \over S}\equiv\frac{\rho_{\rm s}}{S/V}   \simeq
\frac{ (\gamma_s(t) g_s/\pi^2) T^3 0.5\,x^2K_2(x)}
  {g\,4\pi^2/ 90\, T^3}
=\frac{\gamma_s g_s}{g} 0.23 [0.5 x^2K_2(x)]\,.
\eeq
For early times, when $x=m_s/T(t)$ is relatively small, the equilibrium value ($\gamma_s=1$)
can be as large as $s/S\simeq 0.045$. However, at high temperature strangeness is not yet equilibrated
chemically.  For $m_s/T\simeq 0.7$ appropriate for hadronization stage, 
the QGP chemical equilibrium is reached
 when  $s/S\simeq 0.041$. We will see, in section \ref{equQGP},
that this value is fully achieved at LHC, while at RHIC  we expect to be 10-15\%
below chemical equilibrium in QGP. 

To close this discussion, I note that  consideration of the  production of
strangeness at a fixed temperature $T$, and not at a fixed entropy $S$,
is not advisable, as this frees the outcome
from an important experiment related constraint, and it is easy to claim that  
strangeness will  not equilibrate in QGP.  I observe here   that, when the initial temperature 
is reduced by 20\%, the initial entropy content is cut in half. This means that there are half
as many gluons, and the rate of strangeness production by gluon fusion  is cut down by 
a factor 4. For this reason, we will look at how the specific yield $s/S$ evolves, and, along with 
this,  at the  chemical equilibration  achieved in QGP.

\subsection{Strangeness production}\label{prods}
The kinetic evolution of strangeness in
 the local (co-moving) frame of reference, the rate 
of change of strangeness,   is due to production 
and annihilation reactions only:
\begin{eqnarray} 
\frac{1}{V} {{d N_s}\over {d \tau}}=\frac{1}{V}{{d  N_{\bar s}}\over {d \tau} }
=
\frac12 \rho_{\rm g}^2(t)\,\langle\sigma v \rangle_T^{gg\to s\bar s}
+
\rho_{q}(t)\rho_{\bar q}(t)
\langle\sigma  \rangle_T^{q\bar q\to s\bar s}
 \label{qprod}
-
\rho_{s}(t)\,\rho_{\bar{\rm s}}(t)\,
\langle\sigma v\rangle_T^{s\bar s\to gg,q\bar q}.
\end{eqnarray} 
The thermally average cross sections
are: 
\begin{equation}\label{Tsig}
\langle\sigma v_{\rm rel}\rangle_T\equiv
\frac{\int d^3p_1\int d^3p_2 \sigma_{12} v_{12}f(\vec p_1,T)f(\vec p_2,T)}
{\int d^3p_1\int d^3p_2 f(\vec p_1,T)f(\vec p_2,T)}\,.
\end{equation}
$f(\vec p_i,T)$ are the relativistic Boltzmann/J\"uttner
distributions of two colliding particles $i=1,2$ of momentum $p_i$,
characterized by local statistical parameters. 

The temporal evolution of $s/S$, in an expanding plasma, 
is governed by:  
\begin{eqnarray}\label{qprod3a}
 {d\over { d\tau}} {N_s\over S}
&=&
 {A^{gg\to s\bar s}\over (S/V) } 
    \left[\gamma_{\rm g}^2(\tau)-\gamma_{s}^2(\tau)\right] 
+ 
{A^{q\bar q\to s\bar s}\over (S/V) } 
    \left[\gamma_q^2(\tau)-\gamma_{ s}^2(\tau)\right]\,.
\end{eqnarray}
When all $\gamma_i\to 1$, the Boltzmann collision term vanishes, and
 equilibrium has been reached.  Here, we use the invariant rate per unit time and volume, 
$A^{12\to 34}$, by incorporating the  chemical equilibrium densities into the thermally
averaged cross sections:
\begin{equation}
A^{12\to 34}\equiv\frac1{1+\delta_{1,2}}
\gamma_1\gamma_2  \rho_1^\infty\rho_2^\infty 
             \langle \sigma_{s} v_{12} \rangle_T^{12\to 34}.  
\end{equation}
$\delta_{1,2}=1$ for the reacting particles being identical bosons, 
and otherwise, $\delta_{1,2}=0$.  $\gamma_i$ expresses the deviation
from equilibrium of density $\rho_i$. Note also that
the evolution for $s$ and $\bar s$ in proper time of the 
co-moving volume element is identical as both change in pairs.

\subsection{QCD parameters} \label{QCDinput}
We evaluate $A^{gg\to s\bar s}$ and $A^{q\bar q\to s\bar s}$ employing the available
strength of the QCD coupling, and range of accepted strange 
quark masses. 
The known properties of QCD strongly constrain our 
results, however, it turns out that the range
of strange quark masses remains sufficiently wide to impact the results. 
We employ $m_s(\mu=2\,{\rm GeV})=0.10$ GeV which remains uncertain at the level of 25\%.
We compute rate of reactions employing a running strange quark mass working
in two loops, and using as the energy scale the CM-reaction energy 
$\mu\simeq  \sqrt{s}$. Since the running of mass involves
a multiplicative factor, the uncertainty in the mass value discussed above
is  the same for all values of $\mu$. Some simplification is further
achieved by taking, at temperature $T$, the value $\mu\simeq 2\pi T$ which 
is the preferred value of the thermal field theory, and   agrees with
 the value obtained for  the reaction energy in strangeness producing processes. 
This means that we use $m_s(T)=m_s(\mu=2\pi T)$
with $m_s(T=318\,{\rm MeV})=0.1\,$GeV. 
 
The  strength  of  the QCD couping constant 
is today  better understood. We use  as reference value  
 $\alpha_s(\mu=m_{Z^0})=0.118$, 
and evolve the value to applicable energy domain $\mu$ by using two loops. 
The behavior of $\alpha_s(\mu)$ explains why it makes sense to 
use perturbative methods of QCD to describe strangeness production,
a relatively soft process: given the magnitude   $\alpha_s(\mu=m_{Z^0})=0.118$, one can 
quite well run  $\alpha_s$ to the scale of interest, $\mu>1.2$GeV. 
The strength of the interaction 
remains $\alpha_s<0.5$. In passing, we  note that had the strength of $\alpha_s(\mu=m_{Z^0})$
been 15\% greater, strangeness production could not be studied in perturbative 
approach. We can express $\alpha_s(\mu)$   as function of
temperature by the conditions $\alpha_s(T)=\alpha_s(\mu=2\pi T)$. 
This leads to the  expression:
\begin{equation}\label{alfaseq}
\alpha_s(T)\simeq {\alpha_s(T_c)\over 1+C\ln (T/T_c)},\quad T<6 T_c,
\end{equation}
with $C=0.760\pm0.002$, $\alpha_s(T_c)=0.50\pm0.04$ at $T_c=0.16 $ GeV.
We stress that Eq.\,(\ref{alfaseq})  is a parametrization. Only one logarithm
needs to be   used to describe the two loop
running with sufficient precision, since the range we consider
is rather limited, $0.9T_c<T<6T_c$.  

In the presented results for strangeness production,  
we introduce a multiplicative factor $K=1.7$, 
describing the difference between the leading 
and higher order cross sections. In our case 
this $K$-factor accounts for processes odd in power of $\alpha_s$, 
such as gluon fusion into strangeness, with  gluon 
bremsstrahlung emitted by one of the strange quarks. These 
processes cannot be
accounted for in a study of scale dependence of the coupling strength, 
and  strange quark mass. Being distinguishable, `even' and `odd'  terms
are contributing  incoherently, always increasing the production rate. 
The magnitude of $K$, as required here, has not been computed, the rough magnitude is 
estimated based on   perturbative QCD  Drell-Yan  lepton pair production\cite{Hamberg:1990np},
and   heavy quark production \cite{Kidonakis:2004qe}.

\subsection{Achievement of chemical equilibrium in QGP}\label{equQGP}
\begin{figure*}[tb]
\vskip -0.5cm
\psfig{width=7.0cm,figure=\pathnow    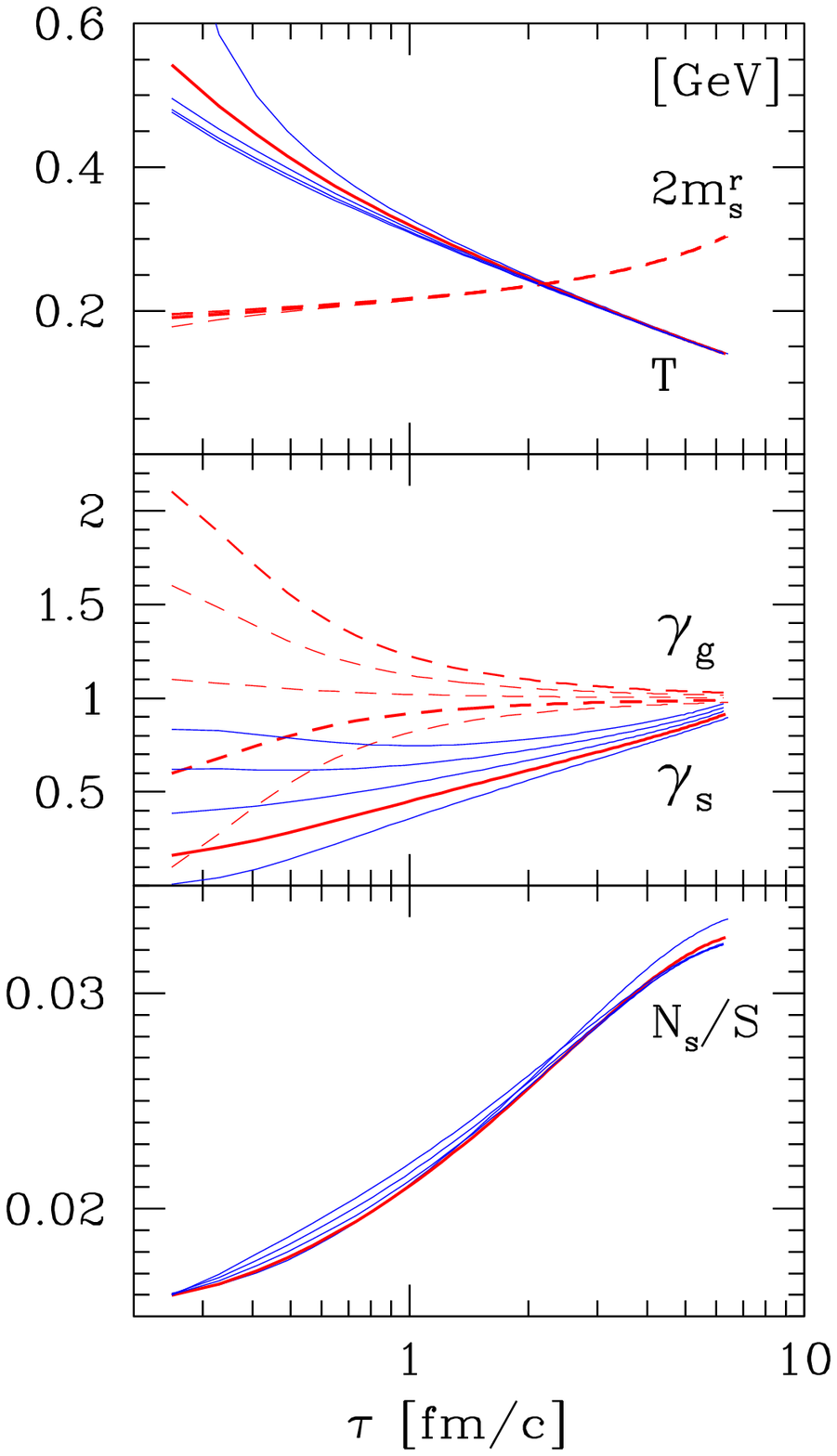  }
\hspace*{-0.6cm}
\psfig{width=7.0cm,figure=\pathnow  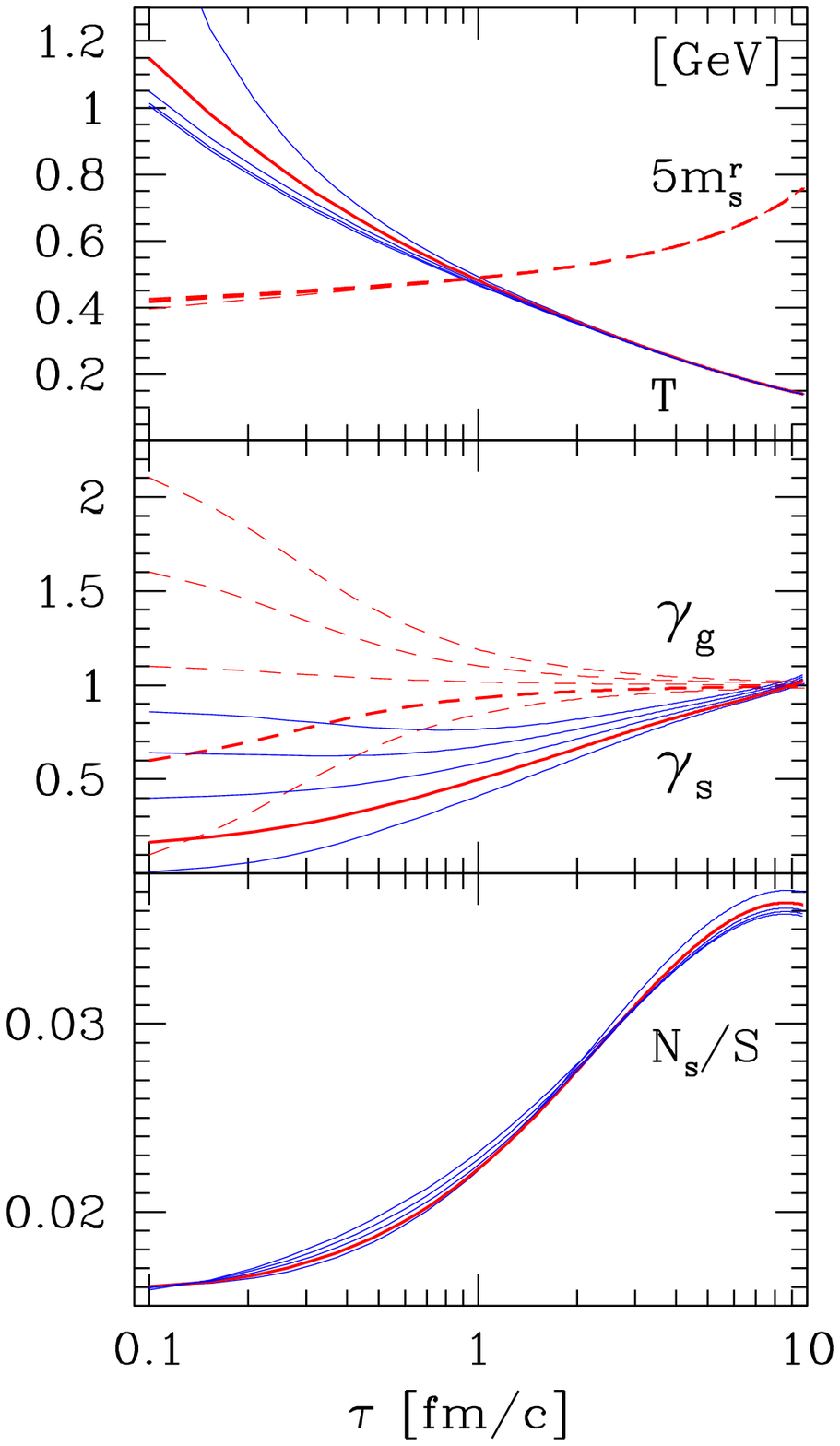   }
\caption{\label{Gluedep}
(color online)  RHIC (left) and LHC (right) strangeness production as function
of fireball proper time. Volume geometric expansion for most central  5\%  collisions is assumed. 
Top panels show the evolution of temperature $T$ and of strange quark mass $m_s(T)$. Middle panel
the assumed variation in gluon $\gamma_{\rm g}$ (dashed lines), allowed greatly different initial conditions. 
  Solid lines, middle panel, show the resulting $\gamma_s$, and in the bottom
panel  the resulting $N_s/S$.
}
\end{figure*}

An important question is how the value of  the
   unknown initial conditions enters. We have studied this in depth 
in several different model approaches. The answer `practically no dependence' 
is best illustrated in the figure \ref{Gluedep}, where 
 we show  using a schematic  volume expansion model  geared to work for 
RHIC on left, and on right for LHC. We explore
 a wide range of initial gluon (and quark) occupancy $\gamma_{\rm g}$, 
which  is shown in the middle panel 
by dashed lines, the initial values we consider for glue occupancy
vary  as $0.1<\gamma_{\rm g}(\tau_0)<2.1 $ in step of 0.5.

With 
$\gamma_{\rm g}(\tau_0)$, we also  vary $\gamma_q(\tau_0)$,
which  following the same  functional temporal 
evolution  starting with 2/3 smaller initial value and evolving 
 1.5 times slower.   Note also that the scale,
in top panel, varies between RHIC and LHC  cases: 
 the dashed lines denote  $2m_s$ on left, and   $5m_s$ on right. 
We use the (initial) value of 
$N_s/S=0.016$ and $dS/dy= 5,000$ for RHIC  and, respectively 
$N_s/S=0.016$ and $dS/dy=20,000$ for LHC. We see   a 
corresponding variation in $T_0$ (top panel, left end of solid lines) 
and $\gamma_s$ (left end of solid lines in middle panel). The final 
results for   $\gamma_s(\tau_f)$ (right end of solid lines in middle panel)  
and $N_s(\tau_f)/S$ (bottom panel) are impressively insensitive
to this rather exorbitant diversity of initial conditions at fixed entropy content.  
The spread in $N_s/S(\tau)$ we see in the bottom panel could be seen as a wide 
line width. We also observe  that  the reaction processes which change yield of strangeness 
can compete with the fast $v_\bot>0.5c$ expansion of   QGP only for 
$T>220$ MeV, for lower temperatures the strange quark yields practically 
do not change (strange quark chemical freeze-out temperature in QGP).

We learn  that strangeness cannot probe the very initial QGP conditions  near
$\tau_0$ ---  the memory of the initial history of the reaction is lost, the system
is opaque  for $\tau<2$--3fm/$c$ to the strangeness signature. On the other hand,
and most importantly, for the study here undertaken, 
this also means that the
experimental observables emerging in the break-up 
of the fireball are characteristic of the properties of nearly
chemically equilibrated QGP phase, with some over-saturation 
expected at LHC due to the continued expansion of QGP 
after freeze-out of strangeness yield, as is seen on right 
in \rf{Gluedep}, where values $N_s/S\simeq 0.4> N_s/S|_{\rm eq}$ are seen.
A very interesting aspect of the kinetic evaluation of strangeness
production is that given the value of $s/S$ in QGP, we are able to evaluate the magnitude of 
$\gamma_s/\gamma_q$ after hadronization, to be expected at LHC, as we did at
the end of section \ref{entro}.

\subsection{Is strangeness enhancement  evidence for QGP?}
\subsubsection{Overall strangeness}
What is exactly ``strangeness production  enhancement''. To what can we compare? 
in this report strangeness enhancement  refers to a comparison of
properties of two  matter phases.The first step was the insight
 that chemically equilibrated QGP has a greater (specific) strangeness content than 
can be found in the hadron phase space. Responding to the challenge from  J\'ozsef Zim\'anyi
I found  that this high yield is achievable considering QGP based thermal scattering reactions, in which 
thermally equilibrated perturbative QCD quanta collide. On the other hand similar kinetic 
calculations show that one cannot expect chemical equilibrium in HG, that is, unless QGP phase 
acts as catalyst of strangeness formation. 

Even in the present theoretical study one must avoid comparing apples
and oranges. Comparing the two phases I assure that conserved quantities
are the same, such as entropy or baryon number.  Whenever all particles produced can be detected, 
  I  prefer to relate  strangeness enhancement to the number of baryon participants. On the other hand
when as at RHIC there is limited phase space coverage, I must compare strangeness yield 
in a fixed window of (central) rapidity: $ds/dy$
at a given  hadron multiplicity,   thus entropy  $dS/dy$. It is very fortunate 
that even though entropy  is itself    enhanced by   QGP phase formation, strangeness is more 
enhanced.

 At RHIC central rapidity strangeness per entropy (hadron multiplicity) 
$(ds/dy)/(dS/dy)$ ratio is the  observable which can be explored as a function
of reaction energy  dependence (strangeness  excitation function)  and centrality dependence.
However, this observable   is hard to evaluate precisely, many particles need to
be measured in same conditions
 and data analyzed in depth to fill acceptance gaps in kinematic and particle type
domains. Thus to report how big the enhancement is, some effort is needed.
On the other hand an enhancement of $(ds/dy)/(dS/dy)$  by 40\% is a huge effect, very well visible, 
 skewing  the general behavior of particle yields towards strange hadrons.

I think all agree and it is quite evident  
that there are many more strange hadrons made in $AA$ reactions than could be expected in 1980.
For example, at central rapidity for the top RHIC energy, there are as many hyperons as
there are non-strange baryons. In the lower energy domain at SPS, there is 
a significant increase in the K$^+/\pi^+$ ratio, when comparing $NN$ with $AA$ reactions.
A cascade of $NN$ reactions under predicts strangeness  yields in presence of QGP,
and to meet this challenge new physics is introduced if one wants to avoid QGP but  overcome  
 strangeness production blocking in confined
phase by high reaction energy thresholds. After 27 years of study 
$AA$ reactions clearly show the systematic features of strangeness enhancement.
Moreover, since there is aside of total strangeness also multistrange hadrons, we have
 a rich and diverse observable.  

\subsubsection{Comparing $NN$  and $AA$  individual particle yields}
For $NN$ reaction models multistrange 
hadrons remain an insurmountable challenge. The QGP benefits from 
the high strangeness density which helps to form multistrange hadrons 
in the final state. However, in order to be able to speak of enhancement of individual hadrons, such as 
$\phi$ or $\overline\Omega$ it is necessary to generate a prediction of 
what the yield would be in absence of quark-gluon plasma. One way to proceed
is to compare the yields of these particles originating  in chemically equilibrated 
QGP and HG phases. This than is simply $\gamma_s^n$, where $n$ is the 
strangeness content, and   $\gamma_s>1$ is computed such that the high QGP
yield is found in the HG phase . This is the point I made in Ref.\cite{Rafelski:1982ii}.

Experimental groups have come to believe that a better measure of enhancement 
is a  comparison of $AA$  yield to $NN$  or $NB$ reactions 
(where $B<<A$)  at the same nucleon-nucleon reaction energy, and this
measure of experimental enhancement has been widely
accepted. However, this concept is, like the ``theoretical enhancement'' 
I have described above at least in part based on a theoretical  
   model scaling  factor~\cite{Glauber:2006gd,Bialas:1976ed,Antinori:2000ph},
 which describes  how the {\it reference}
yield should increase if the $AA$ collisions were a series of {\em independent} $NN$ 
reactions. The excess over and above this scaling 
is reported to  be due to physics beyond $NN$ interactions.   Thus this ``experimental enhancement'' 
is `gauged' by this scaling model.

While the Glauber-Bialas scaling is proved by experience in many other  environments, there is 
no way I can see how to confirm experimentally this model input for RHI strange  particle production. 
For this reason   when comparing yields in $AA$  with $NN$  reaction  other measures  of enhancement
need to be also thought about.    For example  we can ask  by how much 
the multi-strange hadrons are enhanced   as compared to single strange 
hadrons, since in such double ratio the scaling factor drops out.   In a more 
qualitative statement, one simply observes, that the enhancement of multistrange
hadrons with greater strangeness content is greater. In any case the ``theoretical'' 
enhancement as proposed in 1982 must not be forgotten, since if it is, one 
can fall into a trap, that the enhancement arising in the 
comparison of $AA$  with $NN$  reaction could 
be also argued to arise from $NN$ strangeness  yield suppression.
Let us next see how confusion can arise.

\subsubsection{In search of a transforming discovery}
Many relativistic heavy ion physicists after long years of hard work,
are longing for a big discovery.  They  imagine  that 
the heavy ion collision experiments reached some 
revolutionary form of matter, so new that we are almost blind   to it.  An example 
which challenges such fantasy is that within a factor of two or so the yields of most
hadrons is in chemical equilibrium. This behavior seems extraordinary and many of
my  colleagues are deeply impressed. As I discussed  this  result is confirming 
phase space dominance of hadron production and the  Fermi statistical model.
The phase space dominance picture arises naturally in evaporation of QGP 
in the process we call hadronization.

On the other hand it could be that there are mechanisms leading to phase space
dominance other than QGP. For example it has been suggested   that the yield of hadrons 
in $AA$ reactions is always formed in hadron phase chemical equilibrium,  in analogy to the
Unruh-Hawking effect) radiation  of a small black hole. The Hawking radiation
temperature is inversely proportional to the mass of Black-Hole;
\begin{equation}\label{TBH}
T_{\rm BH}=\frac1 {8\pi G M} = \frac a {2\pi}
\end{equation}
where $G$ is the gravitational constant and $M$ the mass of the black hole, and $a$ the acceleration.
The second form relates the Hawking effect to the Unruh effect.

In order to reach hadronic values of $T$ one must in this picture re-think  confinement as being due,
or simply said, being  an event horizon. In that case, by dimensional analysis and/or some 
wishful thinking~\cite{Castorina:2007eb}:
\begin{equation}\label{GCon}
G^2\propto \frac 1 {M_{\rm Planck}^4}\to \frac 1 {32 \pi m_h^2 \sigma}
\end{equation}
where $m_h$ is hadron scale (GeV) mass, and $\sigma$ is the quark string tension. 
Clearly, I have somehow to assure universality of primary hadronization temperature:
somehow just the right   confinement-holes need to be produced reproducibly. So 
I claim that this is so since the hadron scale $m_h$ is fixed by an extra 
dimension. In that way we could view $ m_h^2 \sigma\to \rho$ where $\rho$ is the quark-stretch 
energy in 1+2 dimensions (three regular space less two transverse dimensions + two extra sub-space 
dimensions). $\rho$ thus  has energy density dimension.  Equipped with the proper factor it 
must have as scale the usual 1 GeV/fm$^3$. 

So summarize, QCD is to be forgotten, there is some unknown dynamics which expresses itself
as Hawking-type radiation emanating from reduced spatial and subspace dimensions.
 In this picture   
particle production is computable given the evaporation temperature of the ``confinement-hole'',
there is no particle production dynamics. Equilibrium is in fact always ``natural''.  However,
temperature of the spectra and yields could be still a function of size/mass scales, that is subspace
dimensionality, and the ideas about chemical freeze-out, thermal freeze-out and matter flow 
will need to be converted into this new language --  black-hole-confinement picture 
reduces dimensionality in normal space
and requires extra dimension to compensate this, replacing phase transition temperature 
by extra dimension structured vacuum properties. Ockham razor edge principle, that is 
simple is beautiful,  says that this is not the right way to proceed.  

\subsubsection{Canonical strangeness suppression}
For this reason let us only retain for the moment the outcome of this, and  postulate
ab-initio chemical equilibrium for all reactions. If that were true,  it is the 
 $AA$ particle yield which is the reference, the variation of particle  production I   
called strangeness enhancement is now  strangeness production suppression in
$NN$ reactions. To make sure nobody misunderstands: in this sub-sub-section  
there is 20 fold suppression of $\Omega$ production in $NN$ reactions, and accordingly 
of other particles~\cite{Tounsi:2002nd}.
 
Naively one would think that this  $NN$  suppression should be explained by a kinetic particle
production model. However this is not fitting the approach with ad-hoc assumption of chemical 
equilibrium for the $AA$ particle yield. It seems on first sight that $NN$ reaction volume 
being small, the chemical equilibrium particle yields are to be computed in canonical rather 
than grand-canonical statistical ensemble. The particle-antiparticle correlation 
implemented within canonical statistical approach~\cite{Rafelski:1980gk}
 produce naturally  a much  reduced yields. However the scale of large volume is
reached when $Rm>1$ which in hadronic world is almost always the case, and thus 
the transition from canonical to grand canonical picture for chemical equilibrium yields 
occurs for volumes of hadron size. Moreover the yields are very sensitive to the choice
of the volume size scale $R$.  This than requires an additional  correlation parameter to smooth 
the $R$ dependence if the ad-hoc assumption of chemical equilibrium is to work~\cite{Kraus:2007hf}.

 Given the high sensitivity to the volume size of canonical suppression in $NN$ reactions
one can  produce the required  enhancement to $AA$ volume, which qualitatively
also describes the hierarchy of single, double and triple strange hadron suppression 
within (large) error. Note that  both $AA$ and $NN$  yields depend sensitively on volume 
and freeze-out temperature, thus there  are at least 
4 relevant parameters (two volumes and two temperatures for $AA$ and $NN$ 
systems), and more refined models  introduce additional correlation length 
parameters. All these can be fine tuned to describe  single, double and triple 
strange hadron suppression and 
overall hadron yield within the considerable data error.
  
However, all those pursuing this type of ad-hoc-chemical-equilibrium  data interpretation 
are asking that the QCD practitioners  forget  about the dynamical particle production work 
of past 40 years.  Instead  we
apply   the hypothesis that some novel, and not yet understood
mechanism   provides a chemical equilibrium yield, and introduce many parameters
to describe the data. Ockham razor edge principle also suggests that this is a wrong approach.
However, in perspective of recent history, what invalidated this `canonical' model  was not its obvious
nonsensical general frame of thought, but the simple fact that something that stands
on the head  (in terms of enhancement becoming suppression) will end having 
a  wrong centrality and energy dependence~\cite{Dainese:2005vk,Becattini:2005xt,Caines:2006vd}.\\
\indent a) {\it centrality} dependence: the experimental data for the enhancement 
show a soft dependence on the reaction volume originating in the slight 
advantage in kinetic-thermal strangeness production 
due to prolonged lifespan and greater initial compression 
for more central  (head-on) reactions. This is  incompatible 
with the high sensitivity, nearly threshold behavior
of the canonical suppression mechanism:\\
\indent b) {\it reaction energy} dependence: the 
experimental enhancement of multistrange hadron production is nearly
the same at top SPS energy and at RHIC.. 
This is so since in the conventional reaction picture both  $NN$ and $AA$ yields rise
modestly, for different reasons but in a comparable measure. However, in the 
canonical model  
 as the reaction energy increases, the $NN$ reaction volume increases in order to absorb all that 
energy into the system (since $T$ is fixed) , and thus 
the micro-canonical suppression  changes rapidly comparing SPS and RHIC. \\

\section{Where we are with Strangeness and QGP}
\subsection{Statistical and sudden hadronization}\label{interact}
\subsubsection{Resonances and Statistical Hadronization}
To make a quantitative SHM model of particle production, we  must deal with strong
interactions  among particles, this is done  by way of  introducing 
hadron resonances, as Hagedorn has proposed \cite{Hagedorn:1965st}.  
While resonances are expression of the ongoing collisions and interactions
between hadrons, we also need to account for hadron resonance yield  when
considering the stable hadron yields. One has to be sure to include all the hadronic resonances
which decay feeding into the  yield considered, {e.g.}, the decay 
 $K^*\to K+\pi$ feeds into $K$ and $\pi$ yields. The contribution
is sensitive to production temperature at which these particles are formed. 

Inclusion of the numerous resonances   constitutes a book 
keeping challenge in study of particle 
multiplicities, since decays are contributing at the 50\% level to 
practically all particle yields, \rf{yieldrs}. A public statistical
hadronization program, SHARE 
has simplified this task considerably \cite{Torrieri:2004zz,Torrieri:2006xi}.

\begin{figure}[tb]
\centerline{
\psfig{width=8.8cm,figure=\pathnow 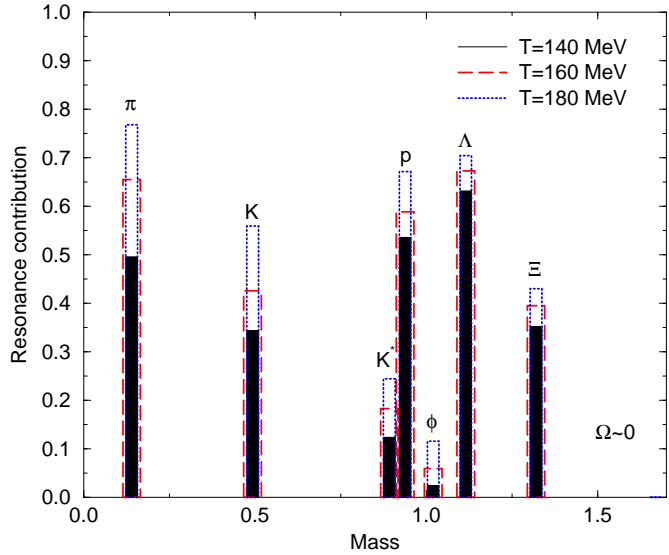}
}
\caption{Relative resonances contribution to the yields of
individual stable hadrons for three particle freeze-out temperatures.
 \label{yieldrs}}
\end{figure}

We see that the  resonance decay contribution is
dominant for the case for the  pion yield. This happens  
even though each pion producing resonance contributes relatively 
little in the final count. However,    the sum of small contributions   competes with the 
direct pion yield. On the other hand, for other, more heavy hadrons, generally
 there is a dominant  contribution from just a few, or even
from a single  resonance. The exception are the $\Omega,\overline\Omega$ 
which have no known low mass resonances: the reason is that the 
ground state has spin 3/2 and the particle being all strange
does not have isospin resonances.

We see that  except for a few particles most stable hadrons are comprising 
the yields of a few main sources which, in general, are of comparable magnitude.
Inclusion of resonances, thus, is a vital element in any proper description of 
a hadronic system. This also implies that we can use resonances to 
test SHM and conversely, without understanding the resonance yields, SHM is
meaningless.

The initial  test  of  statistical hadronization 
approach to particle production is that 
within a particle `family',  particle yields with same valance quark
content are in relation to each other 
thermally  equilibrated. Thus, the relative yield of, {e.g.},
$K^*(\bar s q)$ and $K(\bar s q)$ or $\Delta$ and $N$  is  controlled only
by the particle masses $m_i$,  statistical weights (degeneracy) $g_i$ and the 
hadronization temperature $T$. In the Boltzmann limit,
one has (star denotes the resonance): 
\begin{equation}\label{RRes}
{N^*\over N}= {g^*m^{*\,2}K_2(m^*/T)\over g\,m^{2}K_2(m/T)}.
\end{equation}
Validity of this relation implies insensitivity of the quantum matrix element 
governing the coalescence-fragmentation production of particles to
intrinsic structure (parity, spin, isospin), and  particle mass. 
The measurement of the relative yield of hadron resonances 
is a sensitive test  of the statistical hadronization hypothesis. 

However, the method available to measure resonance yields 
depends in its accuracy on how sudden the hadronization process is.
The observed yield is derived by reconstruction of the invariant mass of the 
resonance from decay products energies $E_i$ and
momenta $p_i$, specifically:
$
m^{*\,2}=(E_1+E_2)^2-(\vec p_1+\vec p_2)^2.
$
Thus,  should  the decay products of resonances rescatter on 
other particles after their formation, 
their energies and momenta will change and normally  the invariant mass
will not fall into the acceptance bin given by the known mass of
the resonance. Hence,  not all produced 
resonances can be, in general, 
reconstructed. 
The rescattering effect depletes more strongly the yields of 
shorter lived states. These decay sooner and are thus more within
the dense matter envelope of the fireball. 

To see how this works, 
 we first evaluate   the ratios of 
$(K^*+\overline{K^*})/K_{\mathrm{S}}$ and
$\Sigma^*(1385)/\Lambda$ using SHM \cite{Rafelski:2001hp,Torrieri:2001ue}.
The upper dashed lines in \rf{ratioT} show the result as function of $T$.
In both cases, chemical potential corrections are negligible since the 
particle's chemical composition is the same:
\begin{equation} 
\label{statistical}
\frac{N^*}{N+N^*} =  \frac{n(m^*,T)}{n(m^*,T)+n(m,T)}, \qquad 
n(m,T) \propto  m^2 T K_2 \left( \frac{m}{T} \right).
\end{equation} 
\begin{figure}[tb]
\centerline{
\epsfig{width=6.7cm,clip=1,figure=\pathnow 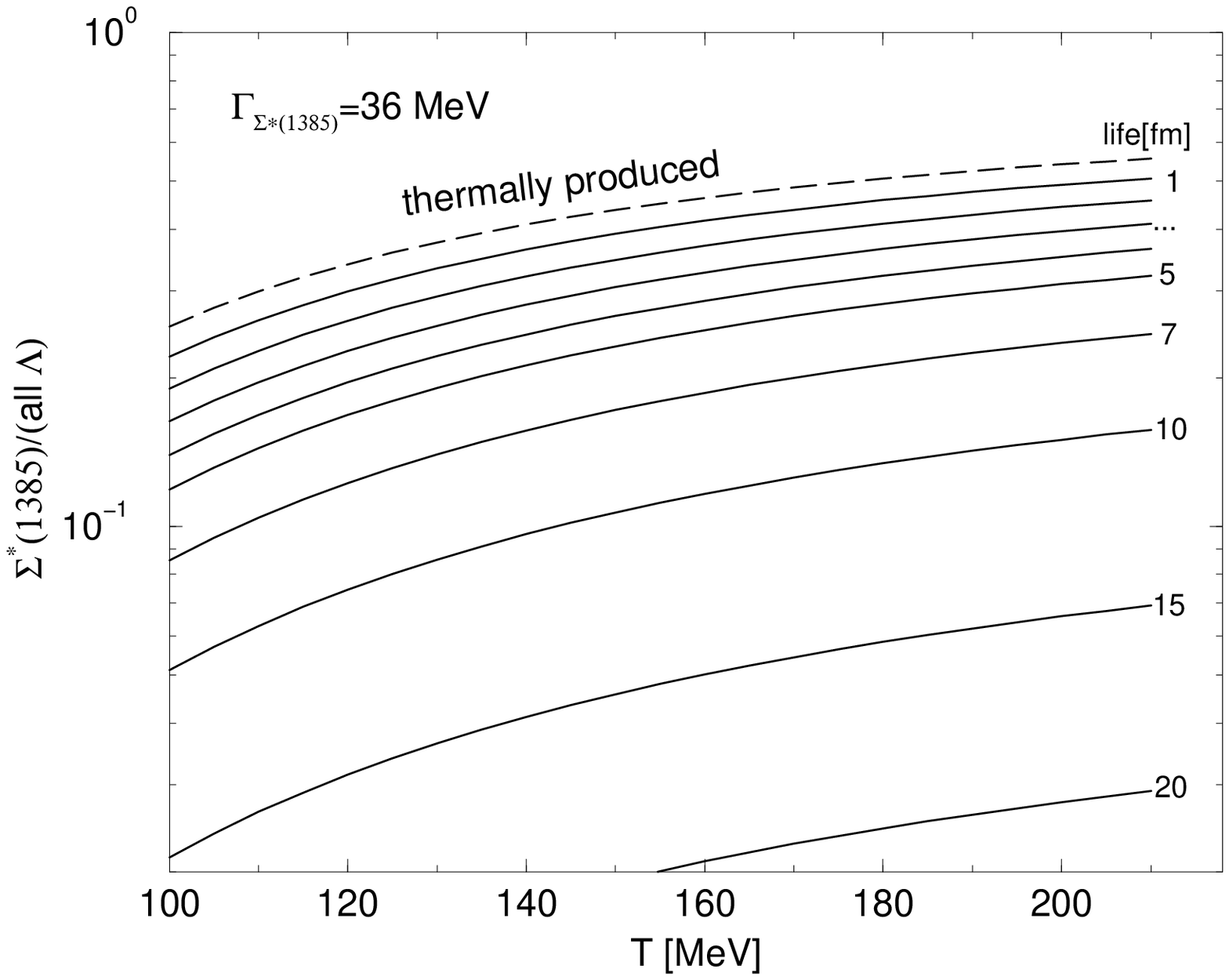}
\epsfig{width=6.5cm,clip=1,figure=\pathnow 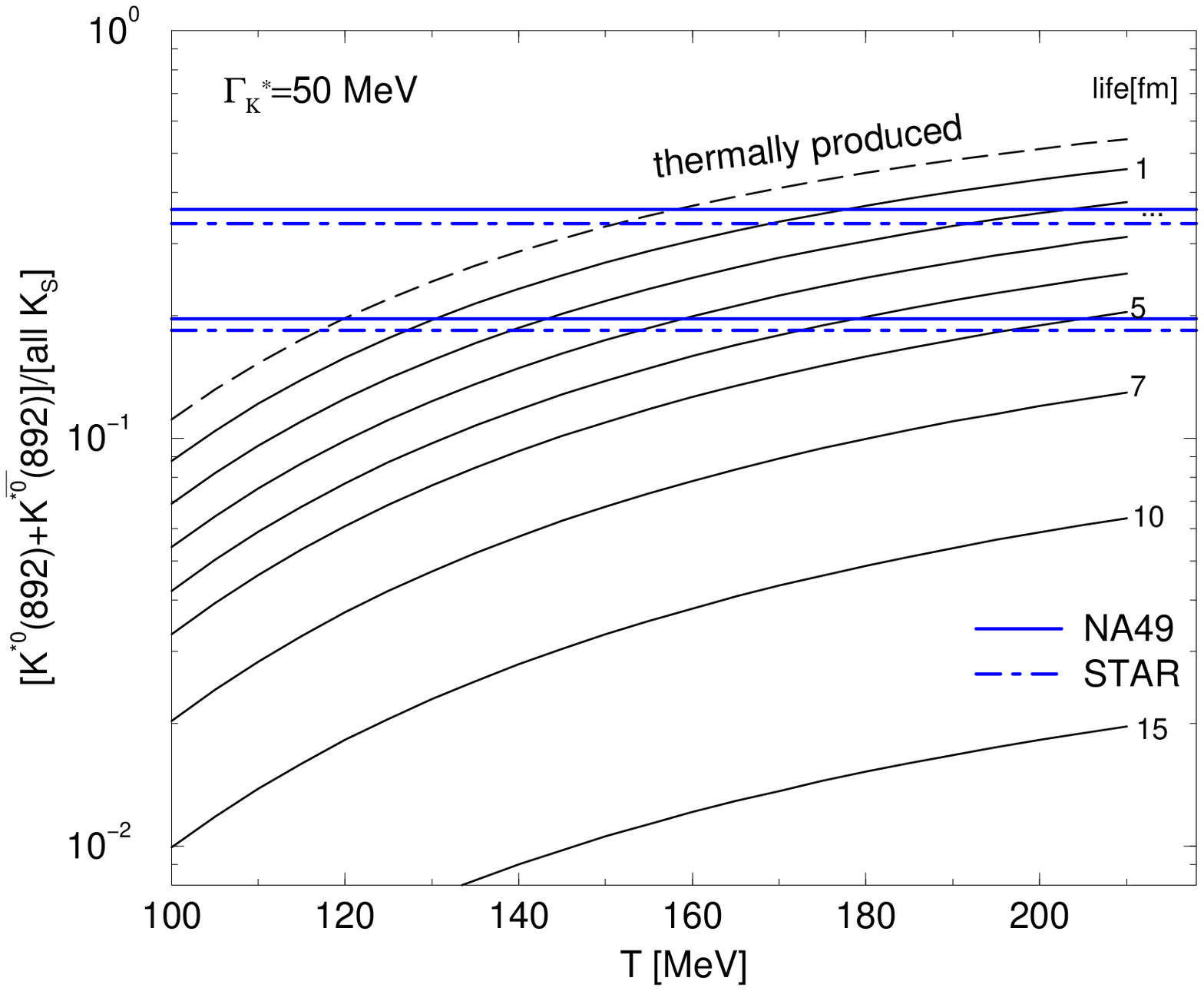}
}
\caption{ Observable relative resonance yields  as a 
function of temperature for a given `life span'
 of  interacting hadron gas phase. 
\label{ratioT}}
\end{figure}

To account for the effect of rescattering, we note that after the resonances decay characterized by
the width $\Gamma$, the decay products undergo rescattering at a rate
proportional to the mediums density as well as the average rescattering rate.
The population time evolution (master) equations then are: 
\begin{eqnarray}
\label{rescattering}
 \frac{d N^*}{d t}&=& -\Gamma N^* + R, 
\quad
 \frac{d D\phantom{^*}}{d t}
=
\Gamma N^* -D\sum_j
\langle \sigma_{Dj} v_{Dj}\rangle \rho_j
          \left(\frac{R_0}{R_0+v t}\right)^3. 
\end{eqnarray}
where the resonance yield is $N^*$ and `daughter' yield is $D$.
$v$ is the expansion velocity, $R_0$ is the hadronization radius, 
$\rho_j=  n_{j} (m_j,T)$ the initial hadron gas particle density
and $\langle \sigma_{Dj} v_{Dj}\rangle$ is the particle specific
thermal  average of  interaction cross-section multiplied with relative
velocity. When $vt\simeq R_0$ the scattering stops, thus qualitatively
$R_0/v\simeq \tau$c is the time of rescattering shown in \rf{ratioT}.

In \req{rescattering},  the regeneration term 
 $R$ is not   relevant   since regeneration reaction rates
are considerably smaller than scattering rates, and as we shall 
see, the predicted lifespan of the system is extremely short. 
However, if and when a long lived system were to emerge,
regeneration could be of relevance \cite{Bleicher:2003ij}.
Figure \ref{ratioT} shows how the observed ratios of 
$\Sigma^*/\Lambda$ and $K^*/K$ are depleted
with varying hadronization time within this model.
The main feature of these results is that a `high' resonance
yield is signaling a short duration of rescattering and thus of 
process of hadronization.  
 
 As indicated in the figure, and more generally, in summary of the many 
experimental results now reported \cite{Markert:2007cg,Adams:2006yu,Salur:2006jq}
the observed yields are relatively large, showing that there is no suppression of the observed
yield for most, if not all, of the resonances observed at RHIC --- a somewhat  `suspect' yield, is
that of $\Lambda(1520)$. It appears  to be low comparing  $\Lambda(1520)/\Lambda$-ratio
seen in $AA$ reactions with those seen in  $pp$ reactions --- and thus, 
we have  considered the possibility that this seemingly stable  $\Lambda(1520)$ state is  in fact
a fragile, easily quenched D-wave,  just like a  quasi-stable $ns$-wave of an atom \cite{Rafelski:2001hp}. 
It turns out that as has been the good tradition, J\'ozsef Zim\'anyi, this time working with 
Peter Levai came across a very similar idea, seeking the effect  in the bound state structure
which is highly unusual \cite{Zimanyi:2004qw}. Potentially, this and other high mass 
negative parity resonances
could be produced in a coalescence process,  with production probability well below the saturation
limit required in the  SHM model. Because of the high mass, their overall  expected yield is sufficiently 
small, so that this suppression will not to alter
the global SHM particle production pattern. 

The story does not end at this point. Namely, a
further explanation we came to realize when working in quantitative fashion with RHIC data is 
in fact, the most natural one: the
fast expansion of the QGP with supercooling. The $AA$  yields are governed, in that case,
 by QGP hadronization temperature which can be noticeably lower than the temperature-like parameter governing  
$pp$ reactions freeze-out, and the reduced relative yield of massive resonances is a signature of
low freeze-out temperature \cite{Torrieri:2006yb}. Indeed, there has been historically a considerable 
theoretical bias  towards supercooled \cite{Csorgo:1994dd},    and thus sudden \cite{Csernai:1995zn},  
hadronization,  beginning with evaluation of recombinant hadron production \cite{Koch:1986ud},  and 
culminating in the usual  Zim\'anyi--Rafelski friendly competition, in the study of 
expanding QGP with confinement considered \cite{Biro:1998dm} leading to supercooling \cite{Rafelski:2000by}.
At this time, the conclusion is that either the low hadronization temperature suppresses the heavy resonances
and thus helps to find the true hadron yields in SHM, or that some of the resonances are suppressed by
hadron structure consideration  in violation of SHM model, in which case their yields do not follow 
the expectations based on \req{RRes}.  Clearly, study of resonance production is of key importance
in understanding heavy-ion collision physics.

\subsubsection{Particle yields and hadrochemistry}\label{ssratios}
In order to find the magnitude of the parameters governing the chemical 
freeze-out, we analyze particle yields and ratios 
in terms of the chemical 
parameters and the temperature. 
Except for direct pions, practically always one can use Boltzmann 
approximation and large reaction  volume, and what follows 
in this subsection assumes that this simple situation applies. 
The chemical factors play often the dominant role in understanding
these yields. 

In order to gain understanding of the chemistry parameters and 
to understand which particle yields are responsible for which results,
it is often appropriate to study (multi) ratios of particle yields as these
can be chosen such that certain physical features can be isolated. For
example, just the two ratios,  
\beql{bLL}
R_\Lambda = 
\frac{\overline{\Lambda}+\overline{\Sigma}^0+\overline{\Sigma}^*+\cdots }{
{\Lambda}+{\Sigma}^0+{\Sigma}^*+\cdots}\simeq \frac{\bar s\bar u\bar d}{sud}=
\lambda_{s}^{-2} (\lambda_{u}\lambda_{d})^{-2} =
e^{2\mu_{\rm S}/T}e^{-2\mu_B/T}\, ,
\eeq
\beql{bXX}
R_\Xi =  
\frac{\overline{\Xi^-}+\overline{\Xi^*}+\cdots }{{\Xi^-}+{\Xi^*}+\cdots}\simeq
\frac{\bar s\bar s\bar d}{ssd}=
 \lambda_{s}^{-4}  (\lambda_{u}\lambda_{d})^{-1}
      =e^{4\mu_{\rm S}/T}e^{-2\mu_B/T} \, ,
\eeq 
lead to a very good estimate of the baryochemical potential and strange
chemical potential, and thus, to predictions of other particle 
ratios \cite{Rafelski:1991rh}.

The   sensitivity of particle yields to phase space occupancy 
factors $\gamma_i$  derives from comparison of hadron yields
with differing $q,s$ quark content, {e.g.}:
\beql{XiLam}
\frac{\Xi^-(dss)}{\Lambda(uds)}\propto 
\frac{\gamma_d\gamma_s^2}{\gamma_u\gamma_d\gamma_s}\
\frac{g_\Xi\lambda_d\lambda_s^2}{g_\Lambda\lambda_u\lambda_d\lambda_s}\,,\qquad
\frac{\overline{\Xi^-}(\bar d\bar s\bar s)}{\overline\Lambda(\bar u\bar d\bar s)}
\propto 
\frac{\gamma_d\gamma_s^2}{\gamma_u\gamma_d\gamma_s}\
\frac{g_\Xi\lambda_d^{-1}\lambda_s^{-2}}{g_\Lambda\lambda_u^{-1}\lambda_d^{-1} \lambda_s^{-1}}\,.
\eeq

In \req{XiLam}, each of the ratios also contain chemical
 potential factors $\lambda_i$. These can be eliminated 
by taking the product of particle
ratio with antiparticle ratio, thus, 
\beql{Xi2Lam}
\frac{\Xi^-(dss)}{\Lambda(uds)}
\frac{\overline{\Xi^-}(\bar d\bar s\bar s)}{\overline\Lambda(\bar u\bar d\bar s)}
=C_{\Xi\Lambda}^2
\left(\frac{\gamma_s}{\gamma_u}\frac{g_\Xi}{g_\Lambda}\right)^2\,,
\quad
\frac{\Lambda(uds)}{p(uud)}
\frac{\overline\Lambda(\bar u\bar d\bar s)}{ \bar p\,(\bar u\bar u\bar d)}
=C_{\Lambda p}^2
\left(\frac{\gamma_s}{\gamma_u}\frac{g_\Lambda}{g_p}\right)^2\,.
\eeq
The proportionality constant $C_{ab}$ 
describes the phase space size ratio for the two 
particles $a,b$ of different mass. It incorporates the contributions
from resonance decays, which of course differ from particle to particle. 

The method applied in \req{Xi2Lam} can be used in several other such double 
particle ratios. The relevance of this is that we have identified an experimental 
observable (combination of particle ratios) 
solely dependent on two parameters of statistical hadronization
and chemical freeze-out, the temperature $T$ which controls the phase space
factor ratio $C$ and the occupancy ratio $\gamma_s/\gamma_q$.
The two  ratios shown in
Eq.\,(\ref{Xi2Lam})),  allow  to constrain the 
value of $\gamma_s/\gamma_q$ only as function of $T$. Other
such   ratios are available. Some, e.g., made of mesons, in general, 
will be weakly dependent on chemical 
potentials since some  kaons and pions 
 are decay products of baryonic  resonances.  For this reason, 
it is more appropriate to study a global fit to the data. However,
considerations as presented here, show better which data is important
and contains the type of information we are seeking to determine. 

Pursuing this further, we note that 
the most difficult to extract, even in a qualitative manner, from the data 
is the value of the parameter $\gamma_q$. One recognizes easily that
the yield of  baryons is proportional to $V \gamma_q^3 (\gamma_s/\gamma_q)^{n_s}$,
where $n_s$ is the total number of constituent  strange quarks or anti-quarks. Similarly, we note that
the yield of mesons is proportional to $V \gamma_q^2 (\gamma_s/\gamma_q)^{n_s}$.
Thus, ratio of baryons to mesons at   fixed value of $\gamma_s/\gamma_q$
is the sole source of information on $ \gamma_q$. What we see is also that 
$ \gamma_q$ allows us to increase the yield of baryons compared to the yield
of mesons. This is a very important feature in statistical hadronization, considering
that microscopic pictures of quark recombination provide yields of baryons and mesons 
which do not  follow the chemical equilibrium distributions, and especially at RHIC 
are quite different such expectations. 

Some   argue that $T$ can fulfill the  function of tuning relative meson to baryon ratio 
as well. However, in 
general, $T$ has to produce the right  resonance   yield and should not
be manipulated to balance baryon to meson yield.  
A consequence of setting $\gamma_q=1$ is that the   overall 
ratio of baryons to mesons fixes the value of $T$ and thus there  ensues a  failure in 
description of resonance yields --- this in turn creates an (as we believe redundant)  industry of 
resonance evolution studies. Despite this effort, there is no consistent 
model which starts with $\gamma_q=1$ and can describe both stable particles and 
resonance yields. Those who pursue this approach must at least 
briefly consider the fact that, in   their  model,  the resonance results remain unexplained. This
in turn means that their use of SHM is totally internally inconsistent 
since the yields of all particles, as we have seen, depend decisively on resonance yields.  
Add to these remarks, the fact that when $\gamma_q\ne 1$ one finds fits of data which
are quite much better than for $\gamma_q=1$.

\subsection{Energy Scan at CERN-SPS}\label{NA49fit}
\subsubsection{Global data analysis}
When facing a large data sample, with many particle yields presented at different
energies, it is better to perform a global fit of statistical parameters, rather than to 
focus on a subset of data such as resonances or hyperon ratios. In that way a 
proper weighting of all errors is arrived at, with the resulting parameter set
applying to all data at the same time.  As already noted the  public statistical
hadronization program, SHARE 
has simplified this task considerably~\cite{Torrieri:2004zz,Torrieri:2006xi}.
As an example of the prevailing situation I will consider here the SPS 
data set of NA49~\cite{NA49Gaz}, the experiment in which  J\'ozsef Zim\'anyi participated. 
The  SPS data is complemented with the highest energy AGS data to show that the 
low energy anomaly we find is present both at SPS lowest energy (20 $A$ GeV) and 
top AGS energy (11 $A$ GeV) , with different experimental set-up and accelerators involved. 
thus the outcome we find is very likely not data but model and analysis method related. 
For the method we refer to  the  detailed presentation of the 
analysis of the top energy AGS results  see Ref.~\cite{Letessier:2004cs}. 

The outcome
of the fit procedure is stated in the top section  of table \ref{AGSPS}.
The $\lambda_s$  values,  marked with
an asterix $^*$ in  table \ref{AGSPS}, are result of a strangeness conservation constraint, 
which, however, is not chosen to be zero, but as shown in table: 
since strangeness conservation constraint involves 
several particle yields it is inappropriate to insist on $s-\bar s =0$, for this 
correlates the errors of the input data, which are experimentally not correlated. Our procedure was to
fit first without strangeness conservation, and once we see the strangeness asymmetry
to fix it at a reasonable nett value shown in table so that there is no spurious constraint
introduced among strange hadrons due to independent measurement, yet there is some 
input of the fact that strangeness is produced in pairs.

\begin{table}
\centering
\caption{For each projectile energy $E$ [$A$\,GeV] for AGS and SPS energy range, we  present in the header  
$\sqrt{s_{\rm NN}}$, the invariant center of momentum
 energy per nucleon pair,   $y_ {\rm CM} $  the center of
momentum rapidity. This is followed
by statistical parameters $T, \lambda_i, \gamma_i$  obtained in the fit, the strangeness asymmetry required,  
and we present the resulting 
 chemical potentials $\mu_{\rm B}, \mu_{\rm S}$,  the reaction volume $V$ and 
the centrality of the reaction considered. This is  followed    first by input and than by output 
total hadron multiplicity $N_{4\pi}$.}\label{AGSPS}
\scriptsize

\begin{tabular}{|c| c | c c c c c |  }
\hline
E[$A$GeV]                      & 11.6          & 20          & 30              & 40         & 80            & 158  \\
$\sqrt{s_{\rm NN}}$  [GeV]     &4.84           &6.26         &7.61             &8.76          &12.32          &17.27  \\
$y_{\rm CM}$                   &1.6  &1.88 &2.08 &2.22 & 2.57& 2.91 \\
\hline
$T$ [MeV]                   &157.8$\pm$0.7  &153.4$\pm$1.6  &123.5$\pm$3    &129.5$\pm$3.4   &136.4$\pm$0.1 &136.4$\pm$0.1       \\
$\lambda_q$                 &5.23$\pm$0.07  &3.49$\pm$0.08  &2.82$\pm$0.08  &2.42$\pm$0.10   &1.94$\pm$0.01 & 1.74$\pm$0.02     \\
$\gamma_q$                 &0.335$\pm$0.006  &0.48$\pm$0.05  &1.66$\pm$0.10  &1.64$\pm$0.04  &1.64$\pm$0.01 &1.64$\pm$0.001\\
$\gamma_s$                  &0.190$\pm$0.009 &0.38$\pm$0.05  &1.84$\pm$0.32  &1.54$\pm$0.15  &1.54$\pm$0.05 & 1.61$\pm$0.02     \\
$\lambda_{I3}$             &0.877$\pm$0.116  &0.863$\pm$0.08 &0.939$\pm$0.023&0.951$\pm$0.008&0.973$\pm$0.002& 0.975$\pm$0.004     \\
\hline
$\lambda_s$                 &1.657$^*$      &1.41$^*$       &1.36$^*$       &1.30$^*$        &1.22$^*$      & 1.16$^*$   \\
$s-\bar s/s+\bar s$    & 0    &-0.092&-0.085 &-0.056&-0.029&-0.062  \\
\hline
$\mu_{\rm B}$ [MeV]               &783      &576            &384            &344             &271           & 227   \\
$\mu_{\rm S}$ [MeV]               &188      &139            & 90.4          &80.8            &63.1          & 55.9  \\
\hline
\hline
$V {\rm [fm}^3]$            &3596$\pm$331   &4519$\pm$261   &1894$\pm$409   &1879$\pm$183    &2102$\pm$53   & 3004$\pm $1  \\
$N_{4\pi}$ centrality          &most central   &  7\%          &  7\%          &  7\%           &  7\%           &  5\%\\
\hline
$R=p/\pi^+$,  $N_W $
                        &$R=1.23 \pm 0.13$  &349$\pm$6      &349$\pm$6      &349$\pm$6      &349$\pm$6      &362$\pm$6           \\
$Q/b $                      &0.39$\pm$0.02  &0.394$\pm$0.02 &0.394$\pm$0.02 &0.394$\pm$0.02 &0.394$\pm$0.02 &0.39$\pm$0.02   \\
$\pi^+$                     &133.7$\pm$9.9  &184.5$\pm$13.6 &239$\pm$17.7   &293$\pm$18     &446$\pm$27     &619$\pm$48       \\
$R=\pi^-$/$\pi^+$, $\pi^-$
                        & $R=1.23 \pm 0.07$ &217.5$\pm$15.6   &275$\pm$19.7 &322$\pm$19     & 474$\pm$28     &639$\pm$48       \\
$R={\rm K}^+\!/{\rm K}^-$,${\rm K}^+$   
                            &$R=5.23\pm0.5$  &40$\pm$2.8     &55.3$\pm$4.4   &59.1$\pm$4.9   &76.9$\pm$6     &103$\pm$10      \\
${\rm K}^-$                 &3.76$\pm$0.47  &10.4$\pm$0.62  &16.1$\pm$1     &19.2$\pm$1.5   &32.4$\pm$2.2   &51.9$\pm$4.9      \\
$R=\phi/{\rm K}^+$, $\phi $ 
                         &$R=0.025\pm 0.006$&1.91$\pm$0.45  &1.65$\pm$0.5   &2.5$\pm$0.25   &4.58$\pm$0.2   & 7.6$\pm$1.1      \\
$\Lambda$                   &18.1$\pm$1.9   &28$\pm$1.5     &41.9$\pm$6.1   &43.0$\pm$5.3   &44.7$\pm$6.0   &44.9$\pm$8.9       \\
$\overline\Lambda$          &0.017$\pm$0.005&0.16$\pm$0.03  &0.50$\pm$0.04  &0.66$\pm$0.1   &2.02$\pm$0.45  &3.68$\pm$0.55    \\
$\Xi^-$                     &               & 1.5$\pm$0.13  & 2.48$\pm$0.19 &2.41$\pm$0.39  &3.8$\pm$0.260  & 4.5$\pm$0.20      \\
$\overline\Xi^+$            &               &               &0.12$\pm$0.06  & 0.13$\pm$0.04 &0.58 $\pm$0.13 & 0.83$\pm$0.04      \\
$\Omega+\overline\Omega$ //  ${\rm K}_{\rm S}$&         &               &               &0.14$\pm$0.07   &              &  81$\pm$4  \\
 \hline\hline
$b\equiv B-\overline B$    & 375.6&347.9 &349.2 &349.9 &350.3 &362.0 \\
$\pi^+$                    & 135.2&181.5 &238.7 &290.0 &424.5 &585.2 \\
$\pi^-$                    & 162.1&218.9 &278.1 &326.0 &461.3 &643.9 \\
${\rm K}^+$                & 17.2 &39.4  &55.2  &56.7  &77.1  &109.7 \\
${\rm K}^-$                & 3.58 &10.4  &15.7  &19.6  &35.1  &54.1  \\
${\rm K}_{\rm S}$          & 10.7 &25.5  &35.5  &37.9  &55.1  &80.2  \\
$\phi $                    & 0.46 &1.86  &2.28  &2.57  &4.63  &7.25  \\
$p $                       &174.6 &161.6 &166.2 &138.8 &138.8 &144.3 \\
$\bar p$                   &0.021 &0.213 &0.68  &0.76  &2.78  &5.46  \\
$\Lambda$                  & 18.2 &29.7  &39.4  &34.9  &42.2  &48.3  \\
$\overline\Lambda$         &0.016 &0.16  &0.51  &0.63  &2.06  &4.03  \\
$\Xi^-$                    & 0.47 &1.37  & 2.44 &2.43  &3.56  &4.49  \\
$\overline\Xi^+$           &0.0026&0.027 &0.089 &0.143  &0.42  &0.82  \\
$\Omega$                   & 0.013&0.068 &0.14  &0.144 & 0.27 &0.38  \\
$\overline\Omega$          &0.0008&0.0086&0.022 &0.030 & 0.083&0.16  \\
${\rm K}^0(892)  $         & 5.42 &13.7 &11.03  &12.4  &18.7  &26.6 \\
$\Delta^{0} $              & 38.7 &33.43 & 25.02 &26.6  &27.2  &28.2  \\
$\Delta^{++} $             & 30.6 &25.62 & 22.22 &24.2  &25.9  &26.9  \\
$\Lambda(1520)$            & 1.36 &2.06  & 1.73 &1.96  &2.62  &2.99  \\
$\Sigma^-(1385)$           & 2.51 &3.99  & 4.08 &4.26  &5.24  &5.98  \\
$\Xi^0(1530) $             & 0.16 & 0.44 & 0.69 &0.73  &1.14  &1.44  \\
$\eta $                    & 8.70 &16.7  & 19.9 &24.1  &38.0  &55.2  \\
$\eta' $                   & 0.44 &1.14  & 1.10 &1.41  &2.52  &3.76  \\
$\rho^0 $                  & 12.0 &19.4  & 14.0 &18.4  &32.1  &42.3  \\
$\omega(782) $             & 6.10 &13.0  & 10.8 &15.7  &27.0  &38.5  \\
$f_0(980)$                 & 0.56 &1.18 & 0.83 &1.27  &2.27  &3.26  \\ 
\hline
\end{tabular}
\end{table}

\begin{figure}[!t]
\centering
\includegraphics[width=9cm,height=13.5cm]{\pathnow 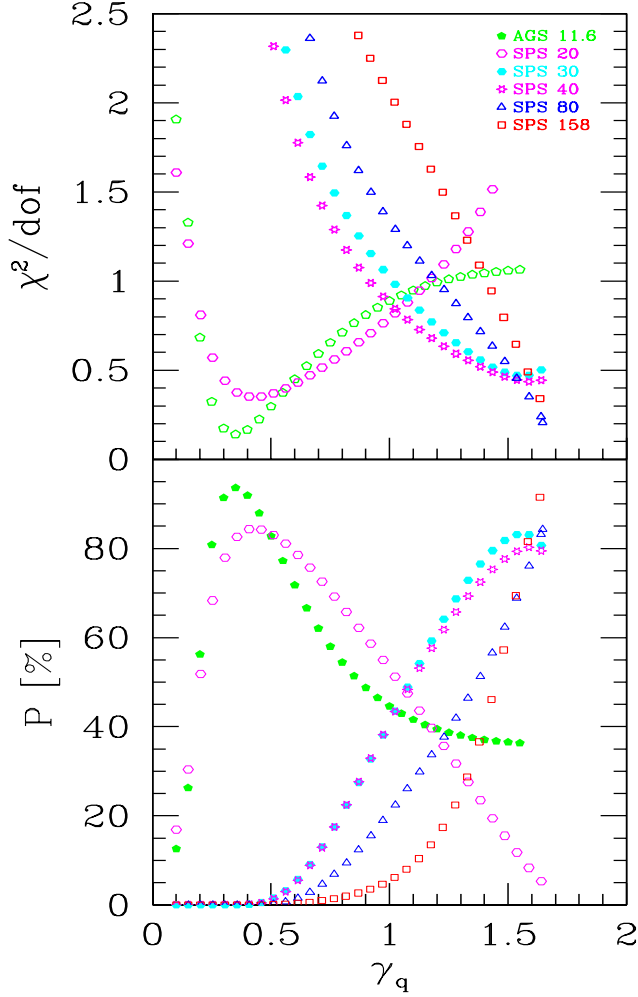}
\caption{\label{ChiP}
$\chi^2/{\rm dof}$ (top) and the associated  confidence level  $P[\%]$
(bottom) as function of $\gamma_q$, the light quark phase space occupancy.
for the AGS/SPS energy range.
}
\end{figure}
 
The  profiles of $\chi^2$, and of the confidence level $P[\%]$ 
determining the fit quality are shown in  \rf{ChiP}.  I note 
that the results for AGS 11.6 and SPS 20 GeV differ from the 
 remainder of the SPS results (30, 40, 80 and 158 GeV) in the outcome of the fit. The low
energy results, obtained at two different experimental  locations, clearly favor a 
value of $\gamma_q<1$, combined with relatively large $V,T$, while the higher energy
data favor  $\gamma_q\to \gamma_q^{\rm cr}\simeq 1.6$.    Inspecting,  in particular, the 80 
and 158 GeV 
profiles, presented in \rf{ChiP}, we recognize that the semi-equilibrium model with 
$\gamma_q=1$  has a comparatively low 
viability compared to the full chemical non-equilibrium model we advance. Thus 
the recent increase in NA49 data sample has tilted the preference strongly to the 
full chemical non-equilibrium interpretation of the data.

 In passing I have to stress 
that  for the present SPS data  these fits are very stable. There is a clear unique best fit.
We trace back this to the fact that the number of participants (both protons and neutrons)
is prescribed by the trigger condition. Leaving this input out generates instability, as does  
ad-hoc approach to charge (isospin) asymmetry - to best of my knowledge only SHARE 
offers at present the correct evaluation of $\lambda3$ parameter which controls charge
asymmetry, and is or particular importance in the study of SPS/AGS data. 

The SHARE package  offers further the opportunity using the statistical parameters to evaluate the physical properties 
of the fireball in its local frame of reference: since we look at the hadron yields,
the flow velocity information is not retained. Some of these results are shown in table \ref{AGSSPSPhysical}.
We note that the chemical freeze-out  at low energy  (AGS 11.6 and SPS 20 GeV) 
occurs from a much more dilute physical state,  the energy density of the high energy  (30, 40, 80 
and 158 GeV) 
data points hoovers well above 400--500 MeV/fm$^3$, about a factor 2.5 higher than at low energy.
Errors on these results can be studied by evaluating a profile of $\chi^2$ varying a fixed value of for 
example $P$. This shows that the errors are at the level of 6\% for extensive quantities $P,\epsilon, S$.
Thus clearly confirms that the difference between these two groups of results is physical. 

I find  that between 20 and 30 GeV  the ratio 
$E/TS$ shifts from a value below unity, to above unity, as is required for
the sudden, supercooled hadronization mechanism expected to operate  for $E>20\ A$ GeV.
There is a steady growth in the yield of strangeness, both measured in terms of $s/S$ as well
as the yield per participant (net baryon number $b$). There is a decrease in the energy retained, indicating that 
the   flow effects grow rapidly, pushing the fraction of energy stopping below 50\% at the top
SPS energy. The cost of strangeness pair production $E_{\rm th}/\bar s$ decreases, as does 
the energy per hadron produced $E_{\rm th}/h$. Both these quantities use energy content in the  local rest frame,
and thus do not include the kinetic energy of  matter  flow at hadronization, which originated from 
the thermal pressure, and which has driven the expansion matter flow. 

\begin{table*}
\centering
\caption{
\label{AGSSPSPhysical}
The physical properties. Top: pressure $P$, energy density $\epsilon=E_{\rm th}/V$, entropy density $S/V$,
  for AGS and CERN energy range at,  
(top line) projectile energy $E$ [GeV]; middle: dimensionless ratios of properties at fireball breakup,
 $E_{\rm th}/TS$; strangeness per entropy $s/S$, strangeness per baryon $s/b$; and bottom
the  fraction of initial collision energy in  thermal
degrees of freedom, $(2E_{\rm th}/b)/\sqrt{s_{\rm NN}}$, the  
energy cost to make strangeness pair $E_{\rm th}/\bar s$,
thermal energy per hadron at hadronization $E_{\rm th}/h$.
}\vspace*{0.2cm}
\begin{center}
\begin{tabular}{|c| c | c c c c c |  }
\hline
E[$A$GeV]                   & 11.6        & 20          & 30          & 40         & 80            & 158  \\
$\sqrt{s_{\rm NN}}$  [GeV]  &4.84         &6.26         &7.61         &8.76        &12.32          &17.27  \\
\hline\hline
$P{\rm [MeV/fm}^3]$         &21.9         &21.3         &58.4         &68.0         &82.3         & 76.9            \\
$\epsilon{\rm [MeV/fm}^3]$  &190.1        &166.3        &429.7        &480.2        &549.9        & 491.8            \\
$S/V{\rm [1/fm}^3]$         &1.25         &1.21         &2.74         &3.07         &3.54         & 3.26            \\
\hline
$E_{\rm th}/TS$             &0.96         &0.92         &1.27         &1.20         &1.14         & 1.11         \\
100$\bar s/S$               &0.788         &1.26     &1.94        &1.90         &2.16          & 2.22       \\
$\bar s/b$                  &0.095         &0.202    &0.289       &0.314        &0.459         & 0.60       \\
 \hline
$(2E_{\rm th}/b)/\sqrt{s_{\rm NN}}$&0.752  & 0.722   &0.612       &0.589        &0.536        & 0.472     \\
$E_{\rm th}/\bar s {\rm\ [GeV]}$   &19.25  &10.9     &8.08        &8.21         &7.19        &6.80  \\
$E_{\rm th}/h {\rm\ [GeV]}$   &1.33  &1.18      &0.866      &0.859         &0.827        & 0.766\\
   \hline
\end{tabular}\vspace*{0.1cm}
 \end{center}
\end{table*}

\subsubsection{Discussion of fit results} 
At the top SPS energy, the value of $s/S=0.022$   implies for a  QGP source a $\gamma_s^Q\simeq 0.7$, which 
corresponds to $\gamma_s/\gamma_q\simeq 1$. This, in fact, is the reason why 
chemical equilibrium $\gamma_q=\gamma_s=1$ `marginally works' 
for this data set.  However, as function of energy we see a very spectacular preference for non-equilibrium,
of two different types. For two lowest reaction energies considered, the results are below 
chemical equilibrium and for other, higher energies, with $\sqrt{s_{\rm NN}}>7.6$ GeV there is
over saturation of chemical occupancies.

The data are fit very well,  and thus also describe precisely the  K$^+/\pi^+$ ratio  as we show in \rf{KPi}. 
This result has been one of major challenges which other groups have failed to understand. The maximum of the 
ratio    K$^+/\pi^+$ occurs  for  $E= 30 \ A$ GeV where I find  $\gamma_i>1$.   An  anomaly associated
with the  horn is the large yield of $\Lambda$, and protons, see bottom section of table \ref{AGSPS}. I further 
note that the structure of the horn  shown by dashed    (semi-equilibrium) and dotted (equilibrium) lines  is also 
reproduced  qualitatively, contrary to results of other 
groups. This  behavior is due  to the relaxation of the strangeness conservation condition, which 
as discussed above cannot be imposed stringently, seen that measurement errors are not correlated. 
\begin{figure} 
\centering
\includegraphics[width=9cm]{\pathnow 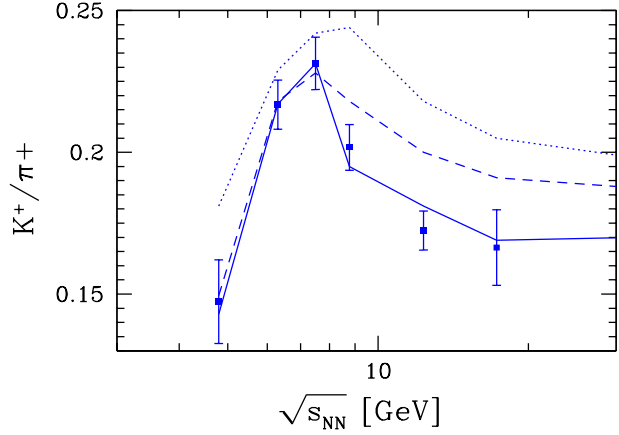}
\vskip -0.2cm
\caption{\label{KPi}
$K^+/\pi^+$ total yields as function of  $\sqrt{s_{\rm NN}}$.
 The solid lines show chemical non-equilibrium  model fit.
 The chemical equilibrium
fit result is shown by the dotted line. The dashed line arises
finding best $\gamma_s$ for $\gamma_q=1$. 
}
\end{figure}
 
It has been argued~\cite{Cleymans:1998fq} 
 that the hadronization condition is defined by the energy available per primary hadron 
$E/h_p$. Using SHARE the  results shown in \rf{Ehp} by triangles arise,  dashed (red) connecting 
lines  indicate results assuming $\gamma_q=1$. The solid (blue)  line connect  results presented
 here, with $\gamma_q\ne 1$. We see that there is a strong dependence
of $E/h_p$ on chemical equilibration conditions at freeze-out: 
$E/h_p(\gamma_q,\gamma_s)$, while the main dependence on $T,\mu_b$ (nearly) cancels,
a feature of considerable interest, but  not at all {\it alone} controlling the hadronization 
process.On the other hand this near cancellation of   $T,\mu_b$  dependence 
is a silent feature allowing to stabilize the search for 
the true physical fit of hadronization data in a highly multidimensional approach which includes 
all chemical parameters. We will return in near future to discuss this in depth~\cite{Petran}.
 It is further interesting to note that the energy cost to produce strangeness seen in \rt{AGSSPSPhysical}
is   higher, roughly by a factor $\pi/$K.
 
\begin{figure}[!t]
\centering
\includegraphics[width=9cm]{\pathnow  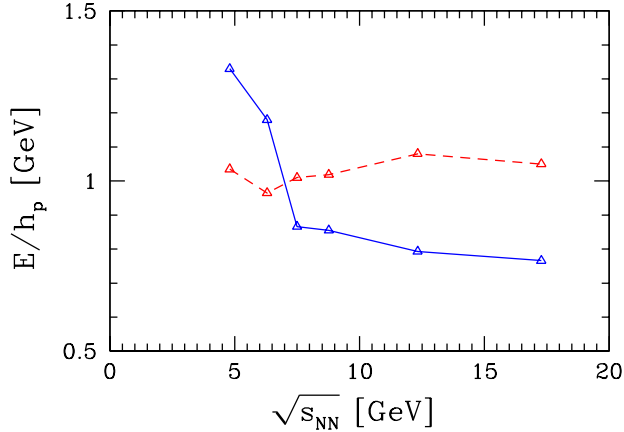}
\caption{\label{Ehp}
$E/h_p$ ratio of the thermal energy of the fireball with total number of primary hadrons,
prior to strong decays as function of reaction energy $\sqrt{s_{\rm NN}}$.  Solid line (blue)
connect results for  chemical
non-equilibrium, dashed line (red) connect results for semi-equilibrium with $\gamma_q=1$.
}
\end{figure}
 
Do the low energy results  imply  absence of quark matter,
and thus reactions between individual hadrons? Our analysis shows
that the chemical freeze-out occurs in a highly dilute phase. However, the rapid rise of strangeness
yield as function of reaction energy,   suggests that the strangeness production
processes differ from those encountered in normal hadron matter.   For this reason we 
favor a constituent quark matter reaction picture at  11.6 and 20 $A$ GeV, with 
color deconfinement arising yet below this energy range. The relatively 
high temperature and low $\gamma_q$ are consistent with properties of  
constituent  quark phase  with $m_{u,d}\simeq 340$ MeV and $m_s\simeq 500$ MeV, 
gluons are `frozen'. In such a massive deconfined quark phase  chiral symmetry is not restored.  
For  $\mu_{\rm B}\to 0$ the lattice results unite the chiral symmetry restoration, in which $m_q\to 0$, with 
the deconfinement transition.

In Summary: the physical properties we find for the hadronization of 30,\,40,\,80,\,158 $A$ GeV
most central heavy ion reactions  correspond to the expected behavior of the chirally symmetric QGP phase. 
SHM model described these results well,  hadron simulations (not discussed here) fail to account for multistrange
(anti)baryons.   11 and 20 $A$ GeV hadronization is different, and importantly the results from AGS and SPS 
lead to the same different results.   What exactly happens, and if, in particular sudden hadronization 
applies in this energy domain is discussed in some more detail
 in another work~\cite{Letessier:2005qe}, and requires  further study. 

\section{Unde venimus, Ubi sumus, Quo vadis}\label{past}
\subsection{From quarks in hadrons to quark-gluon matter}\label{path}
Modern subatomic physics cannot exist  without quarks, yet these particles have never
been seen individually. The introduction 
of a dynamical model of strong interactions~\cite{Fritzsch:1973pi},  created a vast 
new field of study, the structured QCD vacuum. 
Confinement of quarks is interpreted as a feature of the color frozen vacuum 
state in which we live. When in early Universe at about 30$\mu$s, the  ambient temperature dropped
below hadronization condition,  the quark Universe froze and turned into hadrons. 
The early hot Universe did not confine quarks.

The idea to recreate the quark Universe in the laboratory collisions of large nuclei presupposes
that in principle it is possible to embed a small spatial region  of melted, deconfined vacuum 
in the global frozen vacuum. That this is an issue is due to Lorentz/Poincare symmetry, which 
restricts us to a unique vacuum state. However, in
order to heat the vacuum, we crash matter locally, which  breaks the 
Poincare symmetry, allowing to consider two different forms of the Lorentz invariant ``ether''
at the same time.   That this is possible is not entirely  compelling to everyone.
Moreover, there is a time constant of vacuum melting, and I wish to know its value!
If the melting in the conditions of heavy ion collisions is too slow, the  confined state remains. 
Thus formation of QGP is not assured, even though there is a general consensus that it
does not contradict principles of physics, or cumulative knowledge about
the standard model, and hadron structure.

Considering  that a nucleon has a hadronic radius of about $R_N=1$ fm,
and the much greater $u,d$ quark Compton wavelength
$\lambda_{u,d}=h/m_{u,d}c> 100 R_N$,  quark repulsion by the   `frozen' vacuum is required 
to understand hadron structure. The `inside' of a hadron is an excited state of the frozen  vacuum.  
One could easily think this is similar to QGP vacuum. However, 
it is impossible to construct  a model  with   volume energy  above 200 MeV/fm$^3$. 
Yet the study of QCD on the lattice requires that latent heat of 
deconfinement is at the level of 500 MeV/fm$^3$. Thus there must be not one but at least
two different vacuum states aside of quark-empty frozen vacuum around us:\\
a) `half-vacuum' which tolerates quarks, but  no gluons and is found inside the normal hadron;\\
b)  `perturbative thermal QCD vacuum'  which comprises mobile gluons and is called QGP.\\
I still do not quite understand  how to freeze gluons without freezing quarks. Yet, this must be 
possible in presence of relatively high quark density, for not too high temperatures. 

What I want to make sure everyone who read this report to this point recognizes that 
colliding heavy ions can form two thresholds of new phases. 
We must be prepared to see mobility (deconfinement) of 
effectively massive quarks at relatively low excitation energy. Only after quarks turn mass less, 
the chiral  symmetry is restored,  perturbative gluon degrees of freedom should be 
present, and the   quark-gluon plasma phase develops. All this also involves as noted
above the time constant of the vacuum melting which can obscure the two or more phase 
thresholds. Our analysis of data suggests that  this valon (for valance) quark phase  could be   
present at highest AGS energies, and lowest SPS energy range. The valon
semi-relativistic quark matter most resembling squashed nucleons is 
 possibly present  in the center of collapsed stars. 

Such a state of matter  will be  quite different
from the hot QGP state formed with great certainty
in the RHIC and soon LHC environments.  As noted   the latent 
phase heat of this phase could be  different by about a factor 5-10 from QGP-HG latent heat .
Thus there is  need to develop and consider  two different intuitive descriptions of deconfined matter,
one valid at high temperature and low baryon density (QGP) and the other at (relatively) low temperature
and high baryon density (dense quark matter). This alternate quark matter state
 has had a long history~\cite{Barrois:1977xd}, yet only recently there has been 
a surge of  interest~\cite{Alford:2007xm}.

\subsection{Strangeness from AGS to LHC}\label{tools}
In order to study the properties of phases of hadronic matter  formed in relativistic heavy ion 
collisions,  I presented a two step  procedure:\\
a) A model (SHM-SHARE) is developed which allows to {\em reliably} describe the observed particle yields,
and thus to extrapolate  to unobserved phase space domains, or particle types.\\
b) Using all particle yields and the statistical parameters we can compute the physical
properties of the source.  \\
In this way we obtain a snap shot of the hot matter fireball, taken at the time of particle
chemical freeze-out. Much effort in this report was devoted to explain why I chase 
a reliable data fit  with high confidence level: only if this is well accomplished, I have a reliable
extrapolation of the unobserved particle yields, required for precise description of the physical 
properties of the fireball, and I can trust the viability of these results. 

Looking at the snap shot seen in  \rt{AGSSPSPhysical} I note that there is an 
reaction energy threshold which indicates onset of 
the formation of the  perturbative state of QGP   near to 30 $A$ GeV. Interestingly, the rate of growth of
strangeness yield as function of the rate of growth  in the reaction energy diminishes at this point,
which is seen as a knee in some particle yields/ratios.  Since the rate of growth in entropy (pions)
is faster than strangeness, there appears a `horn' structure in K$^+/pi^+$ ratio, shown in \rf{KPi}

Looking at the rapid rise of strangeness yield $\bar s/b$ in \rt{AGSSPSPhysical}  at and below   30 $A$ GeV
I recognize that the  production  of strangeness is rising faster at low reaction energy, in the 
valon-type quark matter which we probably encounter at the top AGS and bottom SPS energy range. 
Among factors helping to explain this  is:\\
a) at lower  reaction energy stopping of matter and energy is more 
effective and thus available  thermal energy   is much greater, which is important considering 
as is seen in   \rt{AGSSPSPhysical} the energy cost of strangeness production. There is a 
rather sudden  drop in stopping between 20 $A$ GeV and  30 $A$ GeV 
(see 3rd last entry line in \rt{AGSSPSPhysical}). \\
b) the more dilute hadronization  condition implies a longer expansion lifespan.\\
c)   the higher temperature profile of this phase -- massive $u,d$-valons  are above threshold for
strangeness-valon production at all times. 

The availability of the yields of multistrange hadrons, and of resonances allows to confirm the 
chemical nonequilibrium (sudden) hadronization picture for the QGP range of reaction
energies (at and above   30 $A$ GeV). Assuming HG phase space chemical equilibrium is simply not a
reasonable model to follow when fitting the data, even if when only looking at the value 
$\gamma_q=1$ in \rf{ChiP} the totality of fits seems to be reasonable there. Only in the
nonequilibrium model can the systematic of (multi)strange hadron production as function 
of reaction energy be described consistently. The horn, \rf{KPi}, shows this, as do similar
results for other particles.I  note  several important roles  the parameter $ \gamma_q$ must fulfill: \\
a) balance entropy  in a fast transition of an entropy rich QGP  to an entropy poor hadron phase;\\ 
b) shift the relative yield of mesons and baryons appropriate for combinant hadronization yields;\\
c) correct for imperfections in the understanding and/or use of hadron spectra.

Regarding the last point I note that it 
cannot  be expected that the empirical baryon and meson spectra which enter the 
SHM are completely understood.   $ \gamma_q$ fudges the incomplete knowledge .
Conversely,  
ad-hock assumption of a value expected in perfect world, viz.   chemical hadron equilibrium  $ \gamma_q=1$ 
presumes  also that the SHM model as used is completely understood. 

It is appropriate to close the discussion noting that 
the theoretical computation of strangeness production based on gluon fusion in QGP
matches analysis results  for  $\bar s/b$
and   $\bar s/S$ seen in  \rt{AGSSPSPhysical}. According to the model calculations 
presented this rise of the production continues 
through the RHIC energy range reaching much higher (nearly double SPS) values of  
$\bar s/S$  at LHC. Thus   yields of strange hadrons at LHC will be further out 
of chemical equilibrium, with relatively large $\gamma_s$ expected. 
For   K$^+/pi^+$ this means that we reach back to, and above the 
peak of the horn at LHC. For multi-strange hyperons, and the $\phi$, 
this implies substantially increased rapidity yields $dN/dy$.

\vspace*{.2cm}
\subsubsection*{Credits}
I wish to thank the students of  J\'ozsef Zim\'anyi: Tam\'as   B\'{\i}r\'o,  Tam\'as Csorgo and 
Peter Levai for  their seminal work addressing the reaction dynamics in heavy ion collisions,  and
quark--gluon plasma which helped to shape my understanding of the subject, and for the very 
stimulating and informative meeting offering a deeper insight about profound 
accomplishments of  J\'ozsef Zim\'anyi.  Comments I received from  Tam\'as   B\'{\i}r\'o,
Peter Levai  and Jean Letessier and an unnamed referee
about this elucidation of hadron  chemistry have greatly improved my presentation and 
the ability to  communicate and explain. 
I thank my friends, students, and  collaborators who published with me   the insights here 
presented, and in particular, (in alphabetical order) Peter Koch, Inga Kuznetsova, 
Jean Letessier, Berndt M\"uller and Giorgio Torrieri.   
\vspace*{.2cm}
\subsubsection*{Support Acknowledgment}
Work supported by a grant from: the U.S. Department of Energy  DE-FG02-04ER4131.
\vspace*{.2cm}
\subsubsection*{Reference Remark}
This is not a review paper of  a vast field of knowledge but a personal account of the work pertinent 
to an important period in life  and work of  J\'ozsef Zim\'anyi. For this reason work most appropriate 
for  this occasion appears below. I apologize to all  colleagues whose work has not been mentioned.

\end{document}